\documentclass[11pt]{article}
\usepackage[preprint]{mlcb_2026}

\usepackage[utf8]{inputenc} 
\usepackage[T1]{fontenc}    
\usepackage{hyperref}       
\usepackage{url}            
\usepackage{booktabs}       
\usepackage{amsfonts}       
\usepackage{nicefrac}       
\usepackage{microtype}      
\usepackage{xcolor}         

\usepackage{graphicx} 
\usepackage{amsmath}
\usepackage{amssymb}
\usepackage{bbm}
\usepackage{url}
\usepackage{multicol}
\usepackage{multirow}
\usepackage{booktabs} 
\usepackage{wrapfig}

\usepackage{environ}

\NewEnviron{hidecontent}{}
\title{Structure-Regularized Interpretable TCR-Epitope Prediction}
\author{%
  Jiarui Li$^1$, Zixiang Yin$^1$, Yunbei Zhang$^1$, Janet Wang$^1$ \\ \textbf{Samuel J. Landry$^2$, Zhengming Ding$^1$, Ramgopal R. Mettu$^1$}\\
  $^1$Department of Computer Science, Tulane University\\
  $^2$Department of Biochemistry and Molecular Biology, Tulane University School of Medicine\\
  \texttt{\{jli78,zyin,yzhang111,swang47,landry,zding1,rmettu\}@tulane.edu} \\
  \textcolor{magenta}{\url{https://github.com/Tulane-Mettu-Landry-Lab/tcr-sr}}
}

\newcommand{\CrossAtten}[2]{\mathcal{A}({#1},{#2})}

\usepackage{xstring}

\newcommand{\ModelName}[1][]{%
    \IfStrEqCase{#1}{%
        {norm}{structure-regularized interpretable model}%
        {cap}{Structure-regularized interpretable model}%
        {first}{Structure Regularized Interpretable Model}%
        {abbr}{TCR-SRIM}%
    }[\textbf{Error: invalid option for \textbackslash ModelName}]%
}

\begin{document}
    \maketitle
    \begin{abstract}
  T cell receptor (TCR)-epitope binding prediction is essential for understanding adaptive immunity and developing immunotherapies. Existing sequence- and structure-based models often generalize poorly to unseen epitopes and provide limited interpretability. Furthermore, the impact of generated structures on model learning remains unclear.
  We present \ModelName[abbr], a structure-regularized interpretable-by-design model that combines protein language model embeddings with interpretable contact prototypes to capture residue-level TCR-epitope interactions. \ModelName[abbr] achieves state-of-the-art predictive performance and improved interpretation quality on the TCR-XAI benchmark. Using its inherent interpretability, we further evaluate the effect of generated structures on model learning. While structures predicted by AlphaFold3, TCRModel2, and tFold-TCR yield competitive performance, they lead to less accurate interaction patterns and reduced binding-site diversity than experimentally-resolved structures. Our results highlight limitations of current structure prediction models for TCR-epitope learning and demonstrate the value of interpretable-by-design models for studying generated biological structures.
\end{abstract}
    \section{Introduction}

In the adaptive immune system, T cells play a pivotal role in recognizing and responding to antigens derived from pathogens, such as viruses and bacteria~\cite{joglekar2021t}, as well as in autoimmune settings. A key step of T cell activation is mediated by the binding between a peptide-Major Histocompatibility Complex (pMHC) and the T cell receptor (TCR). The specificity of this interaction underlies T cell-mediated immunity and is a central focus of therapeutic design and fundamental immunology. A comprehensive understanding of T cell responses is crucial for the development of vaccines and personalized cancer immunotherapies~\cite{rojas2023personalized,poorebrahim2021tcr}.

\begin{wrapfigure}{r}{0.5\textwidth}
    \vspace{-3mm}
    \centering
    \includegraphics[width=\linewidth]{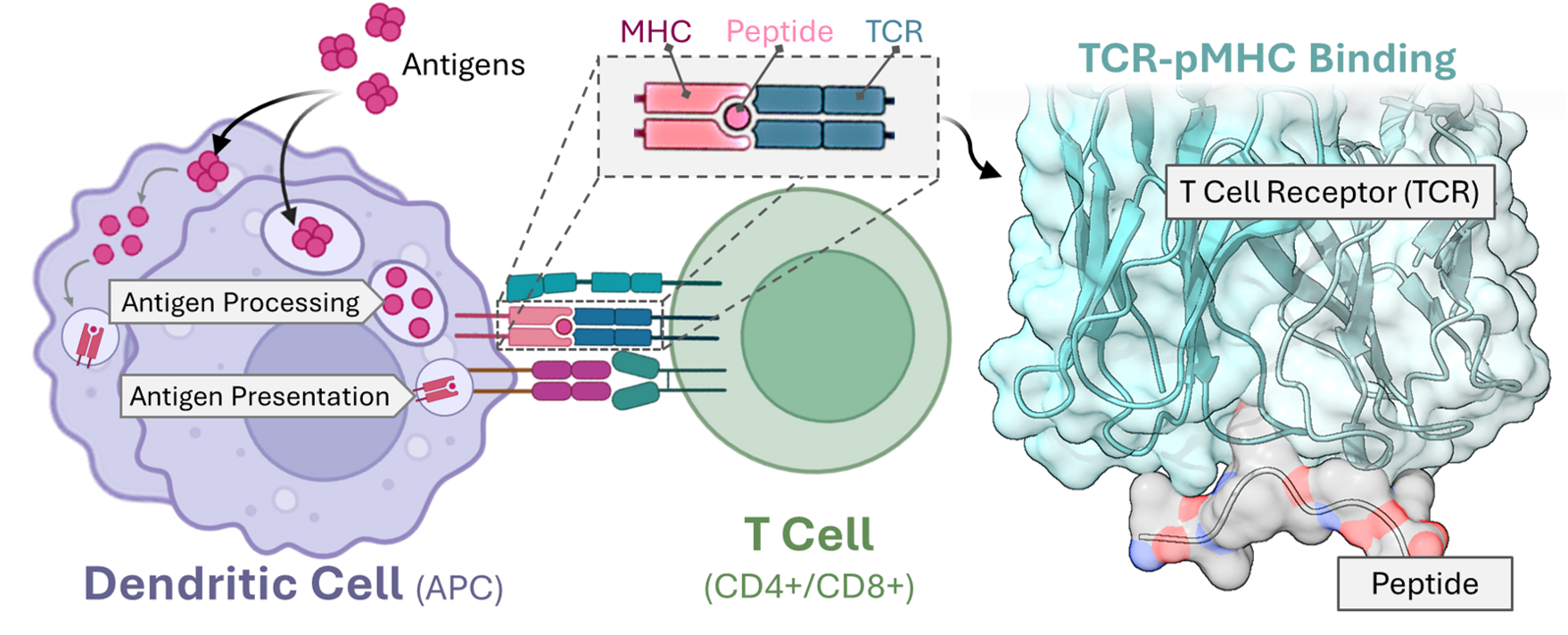}\vspace{-2mm}
    \caption{The interaction between peptide-MHC complexes and T cell receptors is central to adaptive immune responses and critical for the design of vaccines and immunotherapies (figure created in \protect\url{https://www.biorender.com/}).}
    \vspace{-4mm}
    \label{fig:intro:tcrpmhc}
\end{wrapfigure}

As illustrated in Figure~\ref{fig:intro:tcrpmhc}, antigens are first processed by antigen-presenting cells (APCs) and loaded onto MHCI or MHCII molecules, which are then presented on the APC surface as epitopes for recognition by CD8+ or CD4+ T cells, respectively~\cite{davis1988t,neefjes2011towards}. Subsequently, TCRs bind these pMHC complexes, initiating T cell activation. TCR recognition is mediated by the alpha and beta chains, each comprising variable (V), joining (J), and constant (C) regions, with the beta chain additionally containing a diversity (D) region~\cite{bosselut2019t}. Accurate prediction of T cell responses necessitates modeling antigen processing, presentation and TCR-epitope binding~\cite{peters2020t,nielsen2020immunoinformatics}.  

While effective tools for MHCI and MHCII binding/presentation have been developed over the last two decades, the prediction of TCR-pMHC binding remains a central challenge in quantitative immunology and adaptive immune response modeling~\cite{hudson2023can}. Existing approaches are primarily sequence-, structure-based, or a combination of both. Sequence-based methods include both unsupervised and supervised strategies~\cite{hudson2023can,hudson2024comparison}: unsupervised approaches use similarity metrics such as TCRdist3~\cite{mayer2021tcr} on CDRs to cluster TCRs without binding or epitope labels (e.g., GIANA~\cite{zhang2021giana}, GLIPH2~\cite{huang2020analyzing}), whereas supervised models use curated TCR-epitope datasets~\cite{hudson2023can} such as VDJdb~\cite{bagaev2020vdjdb}, McPAS-TCR~\cite{tickotsky2017mcpas}, and IEDB~\cite{vita2019immune}, together with deep learning methods (e.g., MixTCRpred~\cite{croce2024deep}, NetTCR2.2~\cite{jensen2023nettcr}, TULIP~\cite{meynard2024tulip}).

Because experimentally resolved TCR-epitope structures remain scarce in databases such as TCR3D2.0~\cite{lin2025tcr3d} and STCRDab~\cite{leem2018stcrdab}, recent structure-based and sequence-structure hybrid models, including NetTCR-struc~\cite{deleuran2025nettcr} and STAG-LLM~\cite{slone2025stag}, rely on predicted structures from models (e.g., TCRModel2~\cite{yin2023tcrmodel2}, tFold-TCR~\cite{wu2025fast}, and AlphaFold3~\cite{abramson2024accurate}). Despite utilizing structural information, these models still exhibit limited generalization on unseen-epitope benchmarks, while their black-box nature hinders the diagnosis of failure modes and systematic model improvement. To address interpretability, we recently introduced EGM~\cite{li2025rational}, a state-of-the-art TCR-epitope predictor developed with QCAI~\cite{li2025quantifying}, a post-hoc interpretability framework for multimodal TCR-epitope transformers. While EGM shows that interpretability can guide performance gains, it does not incorporate structural information. An alternative is interpretable-by-design models, which embed interpretability directly into their architectures~\cite{rudin2019stop}. Among these, prototype-based networks~\cite{chen2019looks} learn representative patterns that both support prediction and provide faithful interpretations~\cite{chen2019looks,nauta2023pip}. PISTE~\cite{feng2024sliding} adopts this paradigm, but models interactions using predefined biological rules rather than structural information.

To integrate structural information with inherent interpretability, we propose \ModelName[abbr], an interpretable-by-design model that bridges TCR-epitope sequence and structural data for strong generalization and built-in interpretability. \ModelName[abbr] embeds sequences with protein language models including ProteinBERT~\cite{brandes2022proteinbert}, ESM-1b~\cite{rives2021biological}, and ESM-2~\cite{lin2023evolutionary}. It explicitly models residue-level CDR3-peptide interactions via \emph{contact prototypes}, which are regularized with a small set of real or predicted structures while the model is trained on large-scale sequence data.
\ModelName[abbr] outperforms all baselines: paired with ProteinBERT and regularized with real structures, it improves top-100 ROC-AUC by over 9\% relative to TULIP and MixTCRpred, achieves comparable performance on a recent comprehensive TCR-epitope prediction benchmark~\cite{lu2026assessment}, and surpasses existing interpretable methods on our TCR-XAI benchmark by ${\sim}10\%$ in BRHR, a metric rewarding correct attribution to proximal residue contacts~\cite{li2025quantifying}. We further find that predicted structures match real ones in benchmark performance, yet their binding sites diverge from the truth and show homogeneous patterns lacking the diversity of real structures, which limits the generalization of models trained on them. Overall, \ModelName[abbr] achieves state-of-the-art prediction and interpretability while revealing a meaningful gap between predicted and real structures for model learning.

\subsection{Problem Definition}
This section introduces both sequence- and structure-based formulations for the TCR-epitope binding problem.
TCR-epitope binding prediction is a binary classification task: given the CDR3a, CDR3b, and epitope, the model predicts whether they form a binding complex. We omit the MHC molecule, as its identity/sequence is not typically available.
When provided sequences as input, the CDR3a, CDR3b, and epitope peptide are denoted $\alpha \in \mathbb{A}^{n}$, $\beta \in \mathbb{A}^{n}$, and $e \in \mathbb{A}^{n}$, respectively, where $\mathbb{A}$ is the amino acid alphabet and $n$ is the maximum sequence length where inputs are padded as needed. The binding probability is defined as: $p_{\text{bind}} = P(\text{bind} \mid \alpha, \beta, e),$ and a pair is classified as positive if $p_{\text{bind}} > \eta $ with $\eta  \in [0,1]$.

When provided input structures, the CDR3a, CDR3b, and epitope peptide are denoted as $s^a, s^b, s^e \in \Theta$, with $\Theta$ representing the coordinate space for structures of size $n$. Each residue $i$ is represented by the Cartesian coordinates of its atoms, $\theta_i = [(x_{i,j}, y_{i,j}, z_{i,j})]_{j=1}^{n_i}$, where $n_i$ is the atom count of residue $i$. The structural resolution $r \in \mathbb{R}^+$ indicates the quality of the structure, with smaller $r$ corresponding to higher resolution.
    \section{Our Approach}
\begin{figure*}[t]
    \centering
    \includegraphics[width=\linewidth]{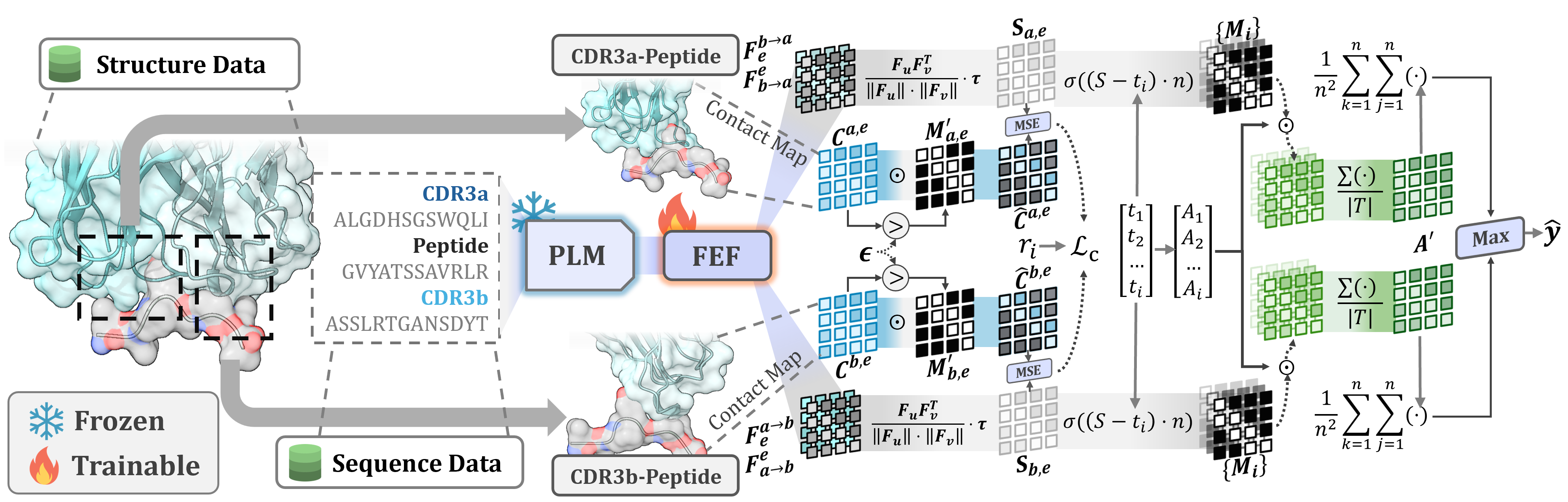}
    \caption{Overview of the proposed \ModelName[abbr] framework for regularizing contact prototypes using structures.}
    \vspace{-0.5cm}
    \label{fig:methods:pipeline}
\end{figure*}

As shown in Figure~\ref{fig:methods:pipeline},
our approach first enhances and fuses features through cross-attention between inputs (i.e., TCR and peptide sequences), then models their contacts using contact prototype layers, and finally regularizes these prototypes with experimentally determined binding structures. The model alternates between standard training on sequence-only data and regularization with structures, thereby incorporating structural knowledge into the prediction model without requiring structures at training or inference.
These components can be directly attached to protein language model (PLM) backbones, which provide embeddings for CDR3a, CDR3b, and peptide sequences, denoted as $\mathbf{E}_a \in \mathbb{R}^{n \times d}$, $\mathbf{E}_b \in \mathbb{R}^{n \times d}$, and $\mathbf{E}_e \in \mathbb{R}^{n \times d}$, where $d$ is the embedding dimension.

\subsection{Protein Language Models}
To obtain richer representations of protein amino acid sequences, several pretrained, self-supervised, transformer-based foundation protein language models (PLMs) have been developed. ProteinBERT is trained on protein sequences and functional annotations, capturing both local and global features for downstream prediction tasks~\cite{brandes2022proteinbert}. ESM-1b is a large-scale transformer pretrained on UniRef50, providing contextualized protein embeddings widely used for structure and function prediction~\cite{rives2021biological}. ESM-2 improves upon this family with larger architectures and expanded pretraining, yielding stronger representations across diverse biological applications~\cite{lin2023evolutionary}.

We extract features from these pre-trained PLMs to ensure our \ModelName[abbr] can work with different PLMs: ESM-1b~\cite{rives2021biological}, ESM-2~\cite{lin2023evolutionary}, and ProteinBERT~\cite{brandes2022proteinbert}. ESM-1b is a 650M parameter model. We used the 8M, 35M, and 650M variants of ESM-2 to investigate how parameter scale affects our \ModelName[abbr] performance. ProteinBERT provides a single 16M parameter model, a PLM backbone that provides both local and global features.

\subsection{Feature Enhancement and Fusion}
In~\cite{li2025rational} we developed an Explanation-Guided Model (EGM), that used post-hoc analyses to inform cross-attention design for TCR-epitope binding prediction. We use the same design for \ModelName[abbr] and denote it as the feature enhancement and fusion module (FEF). For completeness we outline the design below. Formally, we denote cross-attention from arbitrary modalities $q$ to $p$ as $\mathcal{A}_{q\rightarrow p}:\mathbb{R}^{n\times d}\times\mathbb{R}^{n\times d}\rightarrow\mathbb{R}^{n\times d}$ where $q$ serves as the query and $p$ as the key and value. We first obtain cross-fused representations of CDR3a and CDR3b using:
$\mathbf{E}_{a \to b} = \CrossAtten{\mathbf{E}_a}{\mathbf{E}_b}$ and $\mathbf{E}_{b \to a} = \CrossAtten{\mathbf{E}_b}{\mathbf{E}_a}.$
Subsequently, the peptide embeddings are fused further with $\mathbf{E}_{a \to b}$ and $\mathbf{E}_{b \to a}$ to obtain enriched features for TCR-epitope modeling:
\begin{align}\label{eq:fused_embed}
    \mathbf{F}^{a \to b}_{e} = \CrossAtten{\mathbf{E}_e}{\mathbf{E}_{a \to b}},\quad
    \mathbf{F}_{a \to b}^{e} = \CrossAtten{\mathbf{E}_{a \to b}}{\mathbf{E}_e}, \quad
    \mathbf{F}_{e}^{b \to a} = \CrossAtten{\mathbf{E}_e}{\mathbf{E}_{b \to a}},\quad
    \mathbf{F}_{b \to a}^{e} = \CrossAtten{\mathbf{E}_{b \to a}}{\mathbf{E}_e}.
\end{align}

\subsection{Contact Prototype Layers}
Residue-level contacts between TCR and pMHC are a key determinant of binding specificity. TCRdist, a widely used method for TCR-epitope prediction, defines similarity as a weighted mismatch distance between potential pMHC-contacting loops of two receptors~\cite{dash2017quantifiable}. Similarly, PISTE incorporates TCR-epitope contact rules into the attention mechanism to improve both predictive performance and interpretability~\cite{feng2024sliding}. Motivated by these approaches, we design prototype-based layers to explicitly model contacts between TCR and pMHC.

These layers estimate residue contacts between CDR3a and peptide, and between CDR3b and peptide, respectively. For two arbitrary fused embeddings $\mathbf{F}_u\in\mathbb{R}^{n\times d}$ and $\mathbf{F}_v\in\mathbb{R}^{n\times d}$ as defined in Eq. \eqref{eq:fused_embed}, the contact prototype layers take them as inputs and calculate the ``contact area'' between these chains. Inspired by the cross-attention mechanism, we model contact distance through cosine similarity as contact prototypes:
\begin{align}
    \mathbf{S}_{u,v} = \tau \cdot \left(\mathbf{F}_u\mathbf{F}_v^{\top}\right) / \left(\|\mathbf{F}_u\|\|\mathbf{F}_v\|\right) \in [0,1]^{n \times n},
\end{align}
where $\|\cdot\|$ denotes L2 norm and $\tau \in \mathbb{R}^+$ is a learnable temperature parameter controlling the sharpness of the similarity distribution. A user-defined threshold set $T = [t_0, t_1, \ldots, t_{|T|}]$ determines contact maps $\mathbf{M}_i := \sigma((\mathbf{S}_{u,v}-t_i)\cdot n)\in[0,1]^{n\times n}$, with $\sigma(\cdot)$ denoting the sigmoid function. Contacts are aggregated by weighting each threshold with $A = \mathsf{softmax}(T)$ and $A_i\in[0,1]$ represents its $i$-th entry:
\begin{align}
    A' = \left({\textstyle\sum}_{i=1}^{|T|}\sqrt{A_i\mathbf{M}_i}\right) / |T|\in [0,1]^{n \times n}.
\end{align}
The overall contact score between $F_u$ and $F_v$ is:
\begin{align}
    w_{u,v} = \left({\textstyle\sum}_{k=1}^{n}{\textstyle\sum}_{j=1}^{n} A'_{k,j}\right) / n^2 \in [0,1].
\end{align}
Defining $h:\mathbb{R}^{n \times d} \times \mathbb{R}^{n \times d} \to [0,1]$ as the contact prototype function, $h(\mathbf{F}_u, \mathbf{F}_v)=w_{u,v}$, the final TCR-epitope contact score is:
\begin{align}
    \hat{y} = \max \left[h(\mathbf{F}_{e}^{b \to a}, \mathbf{F}_{b}^{a \to e}), h(\mathbf{F}_{e}^{a \to b}, \mathbf{F}_{a}^{b \to e})\right].
\end{align}
We adopt a max operation to reflect that, because empirically, one chain often dominates peptide recognition and single TCR chain can support effective prediction~\cite{shah2026unpaired}.

\subsection{Structure Regularization of Contact Prototype}
To regularize the CDR3-peptide contact map, it is necessary to derive the ground-truth contact map from crystallographic structures. For any given pair of structures $\theta^p, \theta^q$ from the protein structure space $\Theta$, we define the contact between $i$-th residue $\theta^p_i$ in protein $\theta^p$ and the $k$-th residue $\theta^q_k$ in protein $\theta^q$ as $c^{p,q}_{i,k}$, computed as:
\begin{align}
    c^{p,q}_{i,k} = \min_{j=1}^{n_i} \min_{l=1}^{n_i} \sqrt{\left[(x^p_{i,j}-x^q_{k,l})^2 + (y^p_{i,j}-y^q_{k,l})^2 + (z^p_{i,j}-z^q_{k,l})^2\right]/3} \in \mathbb{R}.
\end{align}

The full contact map between protein modalities $p$ and $q$ is then given by $\mathbf{C}^{p,q} = \{c^{p,q}_{i,k}\}_{i=1,k=1}^{n} \in \mathbb{R}^{n \times n}$. Using this approach, the ground-truth contact maps for CDR3a-peptide and CDR3b-peptide interactions are obtained as $\mathbf{C}^{a,e}$ and $\mathbf{C}^{b,e}$, respectively.
The core region of residue-level contacts between TCRs and pMHCs are critical determinants of binding specificity. Motivated by this observation, we utilize the nearest residues between CDR3 regions and the peptide to regularize the model-derived contact prototype $\mathbf{S}_{u,v}$.

For a structural contact map $\mathbf{C}^{p,q}$, we first normalize its scale to align with that of the model-derived contact prototype and invert its values to transform distance measurements into contact importance. The normalized and inverted contact map $\mathbf{\hat{C}}^{p,q}$ is defined as:
\begin{align}
    \mathbf{\hat{C}}^{p,q} = 1 - \left[\left(\mathbf{C}^{p,q} - \min(\mathbf{C}^{p,q})\right) / \left(\max(\mathbf{C}^{p,q}) - \min(\mathbf{C}^{p,q})\right)\right] \in [0,1]^{n\times n}.
\end{align}

We then identify spatially proximal residues by constructing a binary mask $\mathbf{M}_{p,q}' = \mathbbm{1} [\mathbf{\hat{C}}^{p,q} > \epsilon] \in \{0, 1\}^{n\times n}$, where $\epsilon \in [0,1]$ denotes the regularization threshold, which is a hyperparameter.

Then, we regularize the model-derived contact prototype using the top-$\epsilon$ important residues through a masked mean squared error loss:
\begin{align}
    \mathcal{L}^{r} = \left\{{\textstyle\sum}_{i=1}^{n} {\textstyle\sum}_{j=1}^{n} [\mathbf{M}_{p,q}' \odot (\mathbf{S}_{u,v} - \mathbf{\hat{C}}^{p,q})]_{i,j}^2\right\} / \left[{\textstyle\sum}_{i=1}^{n} {\textstyle\sum}_{j=1}^{n} (\mathbf{M}_{p,q}')_{i,j}\right] \in \mathbb{R},
\end{align}
where $\odot$ is element-wise product.
Since structural resolutions vary across different samples, we incorporate resolution-based weighting to balance the regularization loss within each batch. 
Consider a batch of samples indexed by $B \subseteq \mathbb{Z}^+$ containing $|B|$ structures. 
For this batch, the corresponding resolutions are denoted as $R = [r_{i}]_{i \in B} \in \mathbb{R}^{|B|}$, 
and the regularization losses for each structure are represented as $\mathcal{L}^{r}_{B} = [\mathcal{L}^{r}_{i}]_{i \in B} \in \mathbb{R}^{|B|}$.

To incorporate structure quality into training, the resolution-weighted regularization loss ensures that higher-quality structures contribute more to the overall loss, which is defined as:
\begin{align}
    \mathcal{L}^{r'} = {\textstyle\sum}_{i \in B} \left[\left(e^{-R} / {\textstyle\sum}_{j \in B} e^{-r_{j}}\right) \odot \mathcal{L}^{r}_{B}\right]_i \in \mathbb{R},
\end{align}
where lower-resolution values (i.e., higher-quality structures) correspond to higher importance weights to enable model focus on the high resolution samples.
The overall objective function for structure-based regularization is defined as $\mathcal{L} = \mathcal{L}^{r'} + \mathcal{H}_{\text{CE}}(\hat{y}, y)$, where $\mathcal{L}^{r'}$ represents the resolution-weighted structural regularization loss, $\mathcal{H}_{\text{CE}}$ is the cross-entropy classification loss, and $y \in \{0,1\}$ denotes the ground-truth binding label.

The integrated training procedure alternates between standard supervised training and structure-based regularization. Specifically, prior to each normal training epoch, the contact prototype module is regularized using the structural dataset for a number of epochs to incorporate structural constraints.
    \section{Experimental Analysis}
We evaluate our model from three complementary perspectives: predictive performance, interpretation quality, and the impact of predicted versus experimental structures on accuracy and interpretability. Predictive performance and generalization ability are evaluated using appropriate ROC-AUC metrics on both our compiled datasets and a comprehensive benchmark. Interpretation quality is evaluated quantitatively using the binding region hit rate (BRHR)~\cite{li2025quantifying,li2025rational} metric and qualitatively through representative case studies. Since our approach uses 3D structure in the regularization phase, to analyze the contributions of this step we compare the use of experimental structures with those predicted by AlphaFold3, TCRModel2, and tFold-TCR and evaluate performance and interpretation quality of corresponding prototypes to evaluate the impact of predicted structures to prediction models. Finally, to validate the effectiveness of model design choices, we use ablation studies to characterize the impact of structural regularization, mixing strategies, regularization thresholds, and regularization frequency, which are presented in Appendix~\ref{apdx:ablation}.

\subsection{Training and Test Datasets with Unseen Epitopes}
We constructed a TCR-epitope dataset containing 349,716 paired sequences of TCR alpha and beta chains, covering 2,316 unique peptides, 29,581 CDR3a, and 32,578 CDR3b sequences from Homo sapiens and Mus musculus. Of these, 95.7\% are MHC-I and 4.3\% MHC-II. Data were collected from VDJdb~\cite{bagaev2020vdjdb}, McPAS-TCR~\cite{tickotsky2017mcpas}, IEDB~\cite{vita2019immune}, TBAdb~\cite{zhang2020pird}, and 10$\times$ Genomics~\cite{10x2022a}, retaining only entries with valid CDR3a, CDR3b, and peptide sequences.
\begin{wraptable}{r}{0.6\textwidth}
    \vspace{-0.3cm}
    \centering
    \setlength{\tabcolsep}{6pt}
    \caption{Comparison of ROC-AUC scores with the false positive rate restricted to 0.1 across the top-100, top-150, top-200, and top-250 peptides. \ModelName[abbr] yields improvements of approximately 9\%, and 17\% over MixTCRpred and TULIP, and achieves state-of-the-art.}
    \label{tab:roc_auc}
    \begin{tabular}{llllll}
        \toprule
        \multirow{2}{*}{\textbf{PLM Backbone}} & \multirow{2}{*}{\textbf{Model}} & \multicolumn{4}{c}{Top-$k$ ROC-AUC @ FPR$\le$0.1} \\
         & & \textbf{100} & \textbf{150} & \textbf{200} & \textbf{250} \\
        \midrule
        \multicolumn{2}{l}{MixTCRpred~\cite{croce2024deep}} & 0.906 & 0.773 & 0.698 & 0.653  \\
        \multicolumn{2}{l}{TULIP~\cite{meynard2024tulip}}  & 0.821 & 0.706 & 0.648 & 0.613  \\
        \hline
        \multirow{2}{*}{ProteinBERT~\cite{brandes2022proteinbert}}
            & Linear & 0.772 & 0.675 & 0.625 & 0.595 \\
            & \ModelName[abbr] & \textbf{0.989} & \textbf{0.871} & \textbf{0.773} & \textbf{0.713} \\
        \hline
        \multirow{2}{*}{ESM-1b~\cite{rives2021biological}}
            & Linear & 0.900 & 0.795 & 0.716 & 0.668 \\
            & \ModelName[abbr] & \textbf{0.986} & \textbf{0.862} & \textbf{0.766} & \textbf{0.707} \\
        \hline
        \multirow{2}{*}{ESM2-8M~\cite{lin2023evolutionary}}
            & Linear & 0.830 & 0.719 & 0.658 & 0.621 \\
            & \ModelName[abbr] & \textbf{0.953} & \textbf{0.821} & \textbf{0.734} & \textbf{0.682} \\
        \hline
        \multirow{2}{*}{ESM2-35M~\cite{lin2023evolutionary}}
            & Linear & 0.783 & 0.685 & 0.632 & 0.600 \\
            & \ModelName[abbr] & \textbf{0.977} & \textbf{0.838} & \textbf{0.747} & \textbf{0.692} \\
        \hline
        \multirow{2}{*}{ESM2-650M~\cite{lin2023evolutionary}}
            & Linear & 0.876 & 0.762 & 0.690 & 0.647 \\
            & \ModelName[abbr] & \textbf{0.971} & \textbf{0.837} & \textbf{0.746} & \textbf{0.692} \\
        \bottomrule
    \end{tabular}
    \vspace{-1.1cm}
\end{wraptable}
Following other TCR-epitope training protocols (e.g., MixTCRpred~\cite{croce2024deep}, NetTCR-2.2~\cite{jensen2023nettcr}) we generated negative samples by shuffling TCR-epitope pairs (4:1 ratio); we also directly sampled from 10$\times$ Genomics dataset, which contains negative binding data. The dataset was split 95:5 into training and test sets, where the test set (15,503 samples) includes 288 unseen epitopes. To assess generalization to dissimilar epitopes, we sampled evaluation sets with minimal Levenshtein distances between peptides ranging from 1 to 9 following the approach taken by TULIP~\cite{meynard2024tulip}.

\subsection{Epitope-wise ROC-AUC Analysis}
Following TULIP~\cite{meynard2024tulip}, we report epitope-wise ROC-AUC on our compiled dataset. Unlike the commonly reported overall ROC-AUC (aggregated across all epitopes), epitope-wise ROC-AUC evaluates each epitope and its associated TCRs separately, giving a more informative measure of generalization to unseen epitopes, where the aggregate scores for all models are only about 0.55. We also provide an independent benchmark evaluation in the next section. Following established protocols~\cite{nielsen2024lessons}, we constrain the false positive rate to 0.1 per epitope and evaluate on a test set of exclusively unseen peptides; the top-$k$ epitopes by ROC-AUC are reported in Table~\ref{tab:roc_auc}.

We compare against MixTCRpred~\cite{croce2024deep}, TULIP~\cite{meynard2024tulip}, and PLMs with linear classifiers trained on our datasets. We exclude PISTE~\cite{feng2024sliding}, which targets HLA epitope prediction and requires HLA typing unavailable for most of our samples.
\ModelName[abbr] consistently outperforms all baselines, surpassing TULIP and MixTCRpred across every setting and exceeding PLM backbones with linear classifiers by over 10\% on average. With ProteinBERT, it attains an ROC-AUC of 0.989 on the top-100 epitopes, and roughly 9\% and 17\% above MixTCRpred and TULIP, respectively. We also compare against EGM~\cite{li2025rational}, which shares the same design as FEF; even here, \ModelName[abbr] with ProteinBERT improves by 3-5\%. These gains show that structure-based regularization effectively guides the model toward biologically meaningful CDR3-peptide contact patterns, advancing performance in TCR-epitope recognition.

\subsection{Evaluation on Independent Comprehensive Benchmark}
\begin{wraptable}{r}{0.52\textwidth}
    \vspace{-0.6cm}
    \centering
    \caption{Performance comparison of TCR-epitope binding prediction models on the comprehensive benchmark.}
    \label{tab:assess:auprc}
    \begin{tabular}{lll}
    \toprule
    \textbf{Models} & \textbf{AUPRC} & \textbf{ROC-AUC} \\
    \midrule
    \ModelName[abbr] (ProteinBERT)-1R & 0.575 & \textbf{0.571} \\
    \ModelName[abbr] (ESM2-150M) & \textbf{0.587} & 0.564 \\
    PISTE & 0.567 & 0.564 \\
    vibtcr & \textbf{0.588} & 0.563 \\
    MixTCRpred & 0.540 & 0.546 \\
    \bottomrule
    \end{tabular}
    \vspace{-0.3cm}
\end{wraptable}
A large-scale, independent dataset has been proposed to evaluate TCR-epitope binding prediction~\cite{lu2026assessment}, which we refer to as the ``comprehensive'' benchmark throughout this paper for clarity. This dataset provides two evaluation settings for seen epitopes: one using only CDR3b, and another incorporating CDR3b along with additional features. Since our method utilizes both CDR3b and CDR3a as inputs, we adopt the latter setting (CDR3b with additional features) for evaluation~\cite{lu2026assessment}. We directly use the AUPRC (Area Under the Precision-Recall Curve) and ROC-AUC (Area Under the Receiver Operating Characteristic) reported in Lu et al.~\cite{lu2026assessment} for PISTE (PISTE-reftcr)~\cite{feng2024sliding}, vibtcr (vibtcr-AB)~\cite{grazioli2023attentive}, and MixTCRpred~\cite{croce2024deep}.

Table~\ref{tab:assess:auprc} shows that \ModelName[abbr] with a ProteinBERT backbone and one epoch of structure regularization (\ModelName[abbr] (ProteinBERT)-1R) achieves an ROC-AUC of 0.571, establishing state-of-the-art performance. Meanwhile, \ModelName[abbr] with ESM2-150M achieves an AUPRC of 0.587, comparable to vibtcr (0.588) and outperforming all other models. We also evaluate \ModelName[abbr] with other PLM backbones and more models in Appendix~\ref{apdx:compb}.

\begin{wraptable}{r}{0.65\textwidth}
\vspace{-0.6cm}
\centering
    \caption{Comparison of \ModelName[abbr] on BRHR across different protein language model backbones, where $a \to b$ denotes the evaluation of residues in chain $b$ with respect to chain $a$. \ModelName[abbr] shows improvements comparing to baselines, which demonstrates it effectively transforms structure information to sequence-based interpretable models. The interpretation of baseline models are obtained using post-hoc methods.}
    \label{tab:brhr}
    \begin{tabular}{llcccc}
        \toprule
        & \textbf{Embeddings} & \textbf{Peptide} & \textbf{Peptide} & \textbf{CDR3a} & \textbf{CDR3b} \\
         \textbf{Models}  & \quad\quad\textbf{/} & $\downarrow$ & $\downarrow$ & $\downarrow$ & $\downarrow$ \\
         & \textbf{Post-hoc} & \textbf{CDR3a} & \textbf{CDR3b} & \textbf{Peptide} & \textbf{Peptide} \\
        \midrule
        MixTCRpred & AttnLRP & 0.718 & 0.723 & 0.795 & 0.675 \\
        TULIP & QCAI & 0.702 & 0.634 & 0.798 & 0.646 \\
        EGM & QCAI & 0.782 & 0.734 & 0.740 & 0.841 \\
        \hline
        \ModelName[abbr]
        & ProteinBERT & \textit{0.568} & \textbf{0.996} & \textit{0.392} & \textbf{0.855} \\
        \ModelName[abbr]
        & ESM-1b & \textbf{0.833} & \textbf{0.944} & \textbf{0.818} & \textbf{0.852} \\
        \ModelName[abbr]
        & ESM2-8M & \textit{0.604} & \textbf{0.801} & \textbf{0.804} & \textbf{0.816} \\
        \ModelName[abbr]
        & ESM2-35M & \textbf{0.834} & \textbf{0.944} & 0.782 & 0.746 \\
        \ModelName[abbr]
        & ESM2-650M & \textbf{0.961} & \textbf{0.820} & 0.788 & \textbf{0.809} \\
        \bottomrule
    \end{tabular}
    \vspace{-0.3cm}
\end{wraptable}
\subsection{Evaluation of Interpretability on TCR-XAI Benchmark}
To evaluate interpretability for TCR-epitope binding prediction, we constructed the TCR-XAI benchmark~\cite{li2025quantifying} from all currently available experimental TCR-epitope structures in STCRDab~\cite{leem2018stcrdab} and TCR3d 2.0~\cite{lin2025tcr3d}. Retaining only structures with complete TCR $\alpha$/$\beta$ chains, full peptide sequences, intact CDR3 regions, and non-overlapping MHC and peptide chain IDs yields 274 high-resolution TCR-epitope complexes. We randomly split these 1:1 into structure-regularization and evaluation sets. For each evaluation sample, residue-level distances are computed as the minimal atom-wise distance (1) from each CDR residue to the peptide and (2) from each peptide residue to any CDR atom; smaller distances indicate stronger interactions and serve as interpretation ground truth.

We require a metric linking importance scores from an interpretability method to ground-truth binding, and thus construct one correlating the structural proximity of binding residues with importance scores. While proximal residues are not the exclusive drivers of binding, they play a necessary role in mediating TCR recognition; indeed, prior work such as tcrdist3~\cite{mayer2021tcr} uses proximity to define binding fitness. We therefore define the \emph{Binding Region Hit Rate} (BRHR)~\cite{li2025quantifying} to measure how well an interpretation method identifies true binding residues. For a percentile threshold $t$, we take the top-$t$ residues ranked by contact scores and the top-$t$ ranked by peptide-TCR distances; a residue is a \emph{hit} if it appears in both. Following prior work, we use $t=0.25$, a stringent threshold guaranteeing at least one binding residue per instance. BRHR is computed per chain and averaged over all positive samples.

\begin{wrapfigure}[18]{r}{0.645\textwidth}
    \vspace{-0.6cm}
    \centering
    \includegraphics[width=\linewidth]{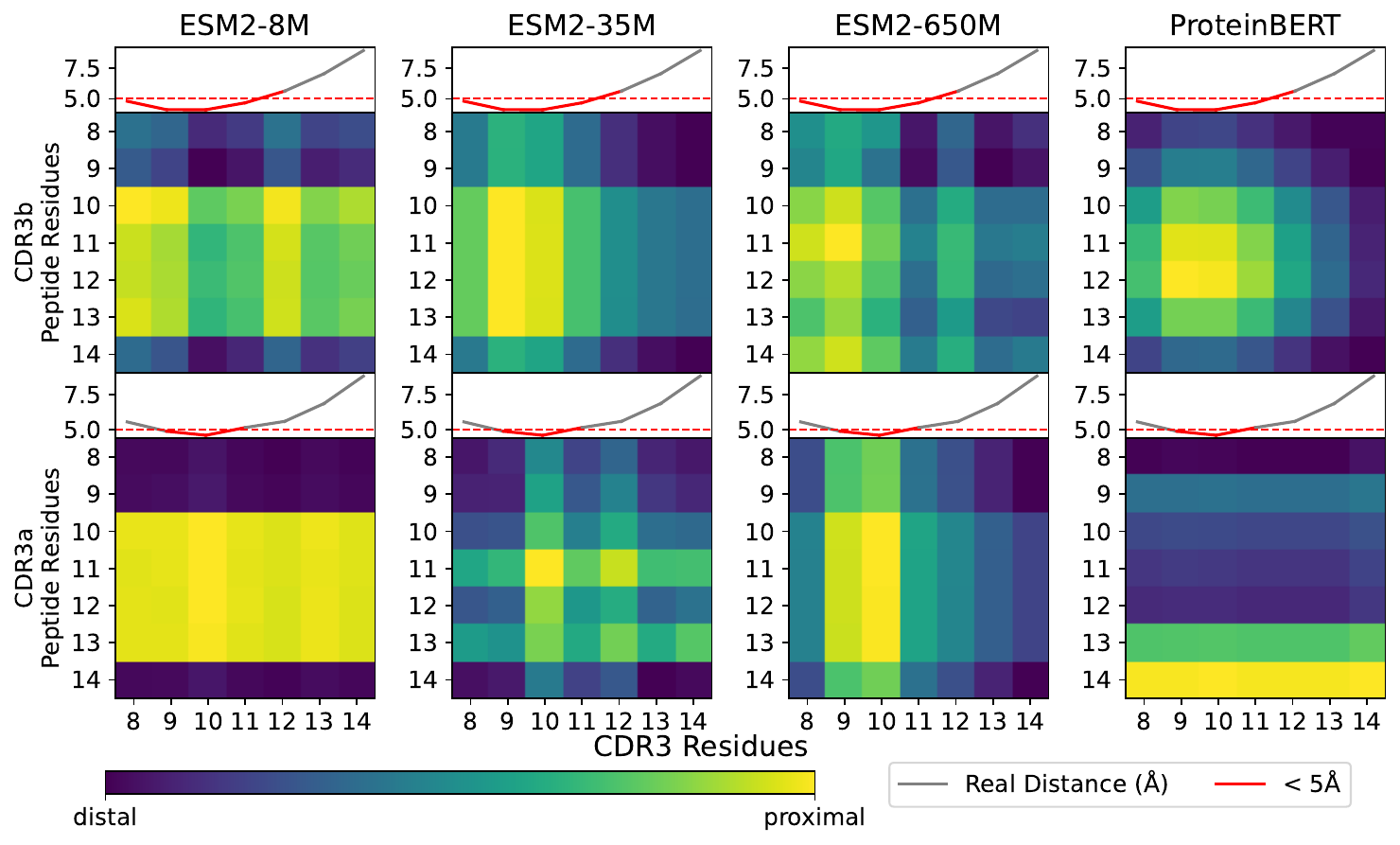}\vspace{-4mm}
    \caption{Visualization of contact prototypes from our \ModelName[abbr] with different PLM backbones. It demonstrates structure regularization effectively transfers contact patterns to sequence-based models.}
    \label{fig:tcrxai:contactmap}
\end{wrapfigure}

As shown in Table~\ref{tab:brhr}, \ModelName[abbr] consistently improves peptide-CDR3b BRHR across all backbones: for peptide $\to$ CDR3b, all models exceed 0.8, with ProteinBERT approaching 100\%. The results also reveal differing backbone strategies. ProteinBERT is near-perfect on CDR3b-peptide interactions but weak on CDR3a-peptide (below 0.4), whereas ESM2-650M reaches ${\sim}0.96$ on peptide $\to$ CDR3a despite moderate CDR3b-peptide performance.

\subsection{Contact Prototype Visualization}
To investigate the interpretations modeled by the contact prototype and their correspondence to true binding across all samples in the TCR-XAI evaluation subset, we extracted and visualized the contact prototypes. For each set of contact prototypes, valid distance values were aligned at the center of the padded contact prototypes and subsequently averaged to reveal the contact patterns captured by the contact prototypes.

As shown in Figure~\ref{fig:tcrxai:contactmap}, we compared the contact prototypes of different parameter variants of ESM2 and ProteinBERT. All models correctly identified the most critical regions of CDR3b, which are closest to the peptide. For CDR3a, the ESM2 series demonstrated notable accuracy, whereas ProteinBERT failed to capture the correct pattern, consistent with the conclusions from our BRHR analysis. Furthermore, for ESM2 models, increasing the number of parameters led to higher ``resolution'' in the contact prototypes, indicating that larger models focus on specific regions rather than producing uniform contact prototypes.
Notably, our \ModelName[abbr] with ProteinBERT produced a highly informative and precise contact prototype, consistent with the BRHR analysis, where it nearly achieved 100\% accuracy. These results show that our method effectively learns the CDR3-peptide contact patterns while providing high-quality interpretability.

\subsection{Case Study}
\begin{figure*}[t]
    \centering
    \includegraphics[width=\linewidth]{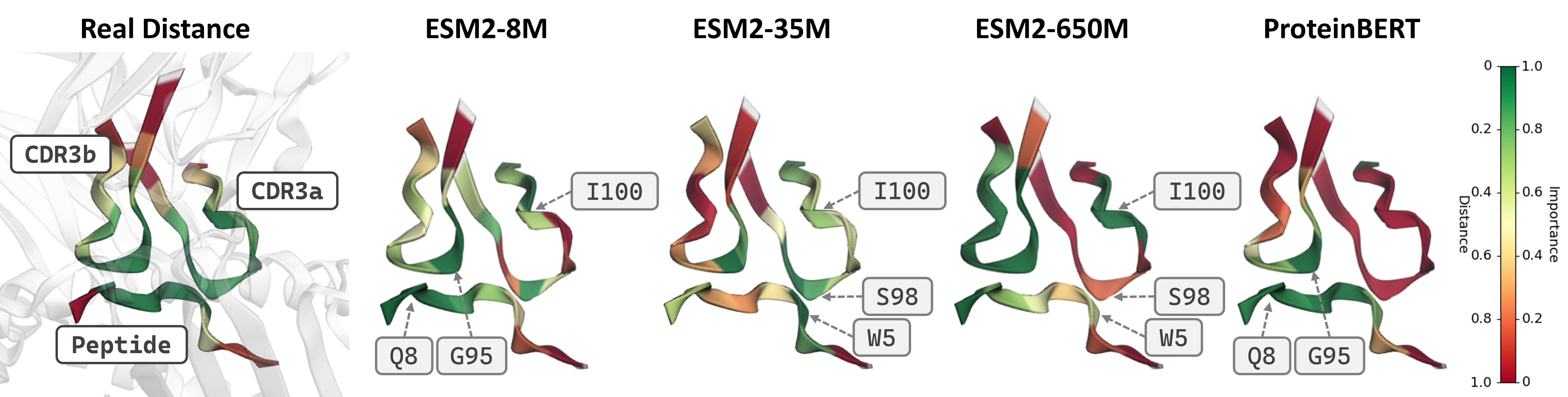}
    \caption{Case study of a tumor-associated peptide analogue with enhanced MHCI binding (\texttt{2BNQ})~\cite{chen2005structural}, illustrating the effects of structure regularization and model-specific contact pattern differences.}
    \vspace{-0.5cm}
    \label{fig:casestudy}
\end{figure*}

To give a concrete example of our approach, we consider a TCR-epitope complex for a tumor-associated peptide analogue with enhanced MHCI binding (PDB: \texttt{2BNQ})~\cite{chen2005structural}, originally investigated to elucidate mechanisms of TCR-epitope binding strength for epitopes in cancer vaccine design.
For this case study, we analyzed the contact prototypes in \ModelName[abbr] with ESM2 and ProteinBERT backbones. To summarize peptide-TCR interactions for peptide, we aggregate contact scores by averaging those of the CDR3 regions.
As shown in Figure~\ref{fig:casestudy}, all models correctly identified the critical CDR3b residue G95. For the peptide residue Q8, which contacts CDR3b, all models assigned relatively high scores. Except for ESM2-35M, which focused on the CDR3a-peptide contacts at W5 and S98, other models emphasized peptide-CDR3b interactions. \ModelName[abbr] with ProteinBERT ignored CDR3a, consistent with our BRHR analysis. Additionally, all ESM2 variants highlighted CDR3a residue I100, near the C-terminus of CDR3a. These examples reveal differences in contact pattern modeling across models for a specific, experimentally verified TCR-epitope complex.

\subsection{Impact of Generated Structures on Model Learning}
Due to the limited availability of experimentally resolved TCR-epitope complex structures, recent structure-aware prediction methods rely on computationally predicted structures to augment structural training data. However, despite incorporating structural information, the performance gains achieved by these approaches remain limited~\cite{lundegaard2008netmhc}. By integrating an interpretable prototype layer and regularizing prototype learning with predicted structural information, \ModelName[abbr] enables a systematic investigation of how generated structures influence model learning and predictive performance. We use ESM2-8M as backbone for \ModelName[abbr] to identify the difference.
To this end, we constructed three predicted-structure variants of the TCR-XAI regulation dataset using AlphaFold3~\cite{abramson2024accurate}, TCRModel2~\cite{yin2023tcrmodel2}, and tFold-TCR~\cite{wu2025fast}. Structures generated by AlphaFold3 and TCRModel2 were obtained through their publicly available prediction servers. For TCRModel2, structures corresponding to \texttt{4Z7W}, \texttt{4OZH}, and \texttt{5KSB} were excluded after three unsuccessful prediction attempts. For tFold-TCR, we deployed a local version to predict structures.

\subsubsection{Impact of Epitope-wise ROC-AUC}
\begin{wraptable}{r}{0.5\textwidth}
    \vspace{-1.2cm}
    \centering
    \caption{Comparison of ROC-AUC scores with the false positive rate restricted to 0.1 across the top-100, 150, 200, 250, and 300 peptides among \ModelName[abbr] regularized with predicted structures.}
    \label{tab:roc_auc:generated_structure}
    \begin{tabular}{llllll}
        \toprule
        \multirow{2}{*}{\textbf{Structures}} & \multicolumn{5}{c}{Top-$k$ ROC-AUC @ FPR$\le$0.1} \\
         & \textbf{100} & \textbf{150} & \textbf{200} & \textbf{250} & \textbf{300} \\
        \midrule
        Real & \textbf{0.953} & \textbf{0.821} & \textbf{0.734} & \textbf{0.682} & \textbf{0.655}\\
        AlphaFold3 & 0.941 & 0.804 & 0.721 & 0.672	& 0.646 \\
        TCRModel2 & 0.929 & 0.790 & 0.711 & 0.663	& 0.638 \\
        tFold-TCR & 0.913 & 0.778 & 0.702 & 0.656 & 0.632 \\
        \bottomrule
    \end{tabular}
    \vspace{-0.4cm}
\end{wraptable}

We first evaluate the generalization performance of \ModelName[abbr] when regularized using experimental versus predicted structures. As shown in Table~\ref{tab:roc_auc:generated_structure}, regularization with real structures yields the best performance, achieving a top-100 ROC-AUC of 0.953. Although regularization with structures predicted by AlphaFold3, TCRModel2, and tFold-TCR results in lower performance, all three variants achieve top-100 ROC-AUC scores above 0.910, where they achieve ROC-AUC scores of 0.941, 0.929, and 0.913, respectively ($p<0.005$ relative to real-structure performance). Notably, this ranking is consistent with the reported accuracy of the structure prediction methods~\cite{wu2025fast}, though we note the predictors may have been trained on structures in the TCR-XAI evaluation set.
These results suggest that predicted structures can serve as effective substitutes when experimentally resolved structures are unavailable, yielding performance comparable to that obtained with real structures. Nevertheless, a consistent performance gap remains, indicating that inaccuracies in generated structures may limit their ability to fully capture the structural information required for optimal TCR-epitope binding prediction.

\subsubsection{Impact of Interpretability}
\begin{wraptable}{r}{0.52\textwidth}
\centering
    \vspace{-1.2cm}
    \caption{Comparison of \ModelName[abbr] on BRHR regularized with different predicted structures.}
    \label{tab:brhr:predstruct}
    \begin{tabular}{lcccc}
        \toprule
         & \textbf{Peptide} & \textbf{Peptide} & \textbf{CDR3a} & \textbf{CDR3b} \\
         \textbf{Structure} & $\downarrow$ & $\downarrow$ & $\downarrow$ & $\downarrow$ \\
         & \textbf{CDR3a} & \textbf{CDR3b} & \textbf{Peptide} & \textbf{Peptide} \\
        \midrule
         Real & \textit{0.604} & \textbf{0.801} & \textbf{0.804} & \textbf{0.816} \\
         AlphaFold3 & \textbf{0.844} & 0.326 & 0.745 & 0.678 \\
         TCRModel2 & \textbf{0.871} & 0.374 & 0.784 & 0.623	\\
         tFold-TCR & \textbf{0.788} & 0.355 & 0.736 & 0.764	\\
        \bottomrule
    \end{tabular}
    \vspace{-0.3cm}
\end{wraptable}
To assess the impact on the choice of regularization structures on interpretability, we evaluate \ModelName[abbr]'s BRHR on the TCR-XAI evaluation set when using experimental versus predicted structures. To do this, we substitute the experimental structures used with those predicted by AlphaFold3, TCRModel2, or tFold-TCR for the same TCR-epitope complex. As shown in Table~\ref{tab:brhr:predstruct}, models regularized with predicted structures yield substantially lower peptide-CDR3b BRHR than the corresponding experimental structure, falling below 0.400 for peptide $\rightarrow$ CDR3b across all predicted variants, while still exceeding 0.780 for peptide $\rightarrow$ CDR3a. Thus predicted-structure regularization preferentially captures peptide-CDR3a patterns while inadequately modeling peptide-CDR3b interactions. Given CDR3b's central role in antigen recognition, this deficiency may explain the performance gap relative to real-structure regularization.
\begin{wrapfigure}{r}{0.645\textwidth}
\vspace{-0.6cm} 
    \centering
    \includegraphics[width=\linewidth]{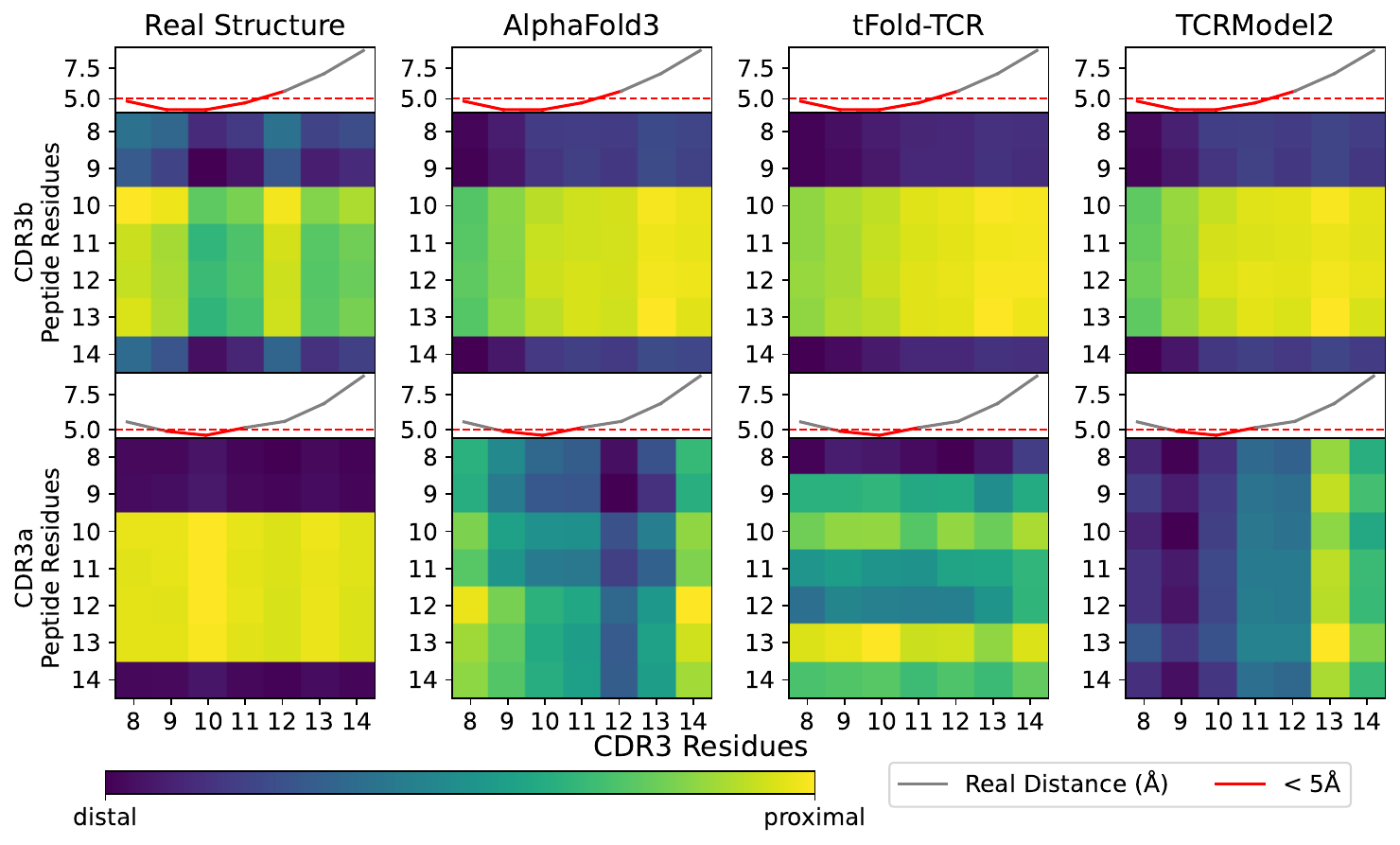}
    \caption{\vspace{-0.3cm}The contact prototypes regularized with predicted structures.}
    \label{fig:cm:pred_structure}
\vspace{-0.4cm} 
\end{wrapfigure}

To characterize the effects of structural regularization, we visualize the contact maps learned by prototypes regularized with different structure sources (Figure~\ref{fig:cm:pred_structure}). For peptide-CDR3a prototypes, only models regularized with tFold-TCR and experimentally resolved structures assign high importance to residue 10, matching the ground truth computed from real structures; AlphaFold3 and TCRModel2 prototypes focus elsewhere and miss it.
For peptide-CDR3b prototypes, predicted structures produce two notable artifacts: shifted binding-site localization and overly smooth contact distributions. Their highest-intensity regions concentrate around residues 13-14, whereas experimentally resolved structures focus on residues 8, 9, and 12, which are more consistent with the observed contact-core spanning residues 8-12. This shift offers an explanation for the reduced peptide-CDR3b BRHR under predicted-structure regularization. The smoother contact maps further indicate lower spatial variance and less localized signals, suggesting that structural prediction errors introduce an over-smoothing effect that captures generic patterns while losing residue-specific contacts.
To probe this, we analyzed the structural variability of CDR loops. For each structure, we computed the CDR-peptide distance matrix, then the pairwise similarity between these matrices as a measure of binding-site diversity, where higher similarity indicates more varied contact patterns. Real structures show significantly similar ($p < 0.005$) than most predicted datasets (1.00 for real CDR3b versus 0.70-0.72 for predicted CDR3b), the sole exception being tFold-TCR on CDR3a (0.99; full results in Appendix~\ref{apdx:sdiver}). Structure prediction methods tend to generate similar CDR conformations despite these regions' inherent flexibility, losing the binding-mode diversity of real complexes. This potentially limits the capacity to learn varied interaction patterns, possibly leading to the weaker generalization under predicted-structure regularization.
    
\section{Conclusion}
\vspace{-0.2cm}
In summary, we present \ModelName[abbr] (\ModelName[norm]), an interpretable framework that bridges sequence- and structure-based modeling through contact prototypes while requiring only sequence inputs at inference. Using PLMs and a small set of high-resolution crystal structures for contact-prototype regularization, \ModelName[abbr] learns CDR3-peptide contact patterns from sequences alone and outperforms state-of-the-art models, showing that structural regularization effectively transfers structural knowledge into sequence models to improve both prediction and, perhaps more importantly, interpretability.
Beyond prediction, \ModelName[abbr] offers a novel framework for analyzing how generated structures shape model learning. Although structures from AlphaFold3, TCRModel2, and tFold-TCR yield competitive performance, they show reduced binding-site diversity and less accurate peptide-CDR3b interaction patterns than experimentally resolved structures, highlighting a key limitation of current structure-prediction models for TCR-epitope learning and the value of interpretable-by-design approaches for evaluating predicted structures.

\textbf{Acknowledgments:} Harold L. and Heather E. Jurist Center of Excellence for Artificial Intelligence at Tulane University.
    \newpage

\bibliographystyle{plain}
\bibliography{reference}
    \newpage
    \setcounter{page}{1}
\appendix
\makeatletter
{
    \vskip 0.1in
    \@toptitlebar
    \centering
    {\LARGE
        {\bfseries\@title} \\
        {\Large \bfseries (Supplementary)}
        \par
    }
    \@bottomtitlebar
    \begin{tabular}[t]{c}\bf\rule{\z@}{24\p@}\@author\end{tabular}%
    \vskip 0.3in \@minus 0.1in
}
\makeatother
\section{Performance Evaluation}
In this section, we introduce the baseline models compared to our model. Then, we extend the epitope-wise ROC-AUC and other metrics analysis and comprehensive benchmark to compare additional \ModelName[abbr] variants against other models, and we evaluate \ModelName[abbr] on the IMMREP23 benchmark.

\subsection{Baseline Models}
We consider two categories of comparable models in our evaluations. First, to evaluate PLM-based models, we constructed a standard linear classifier for PLM features. Specifically, we added two fully connected layers with hidden dimension three times the feature dimension and ReLU activation. Each classifier takes concatenated global representations of CDR3a, CDR3b, and the peptide as input and outputs a prediction score. For ProteinBERT, we used the provided global features, and for the ESM models, we averaged local residue-level features to obtain global representations. Second, we compared our models with two recent transformer-based TCR-epitope prediction methods. MixTCRpred~\cite{croce2024deep}, one of the most widely used TCR-epitope models, utilizes all CDR regions as input. TULIP~\cite{meynard2024tulip} is another recent model that outperforms the widely used NetTCR-2.2~\cite{jensen2023nettcr} baseline in terms of accuracy and generalization ability. EGM~\cite{li2025rational} is a TCR-epitope prediction model designed with post-hoc analysis guidance. We utilized EGM2 architecture, but change the inputs from full TCR to only CDR3 regions.

\subsection{Extended Epitope-wise ROC-AUC Analysis}
\begin{figure*}[ht]
    \centering
    \includegraphics[width=\linewidth]{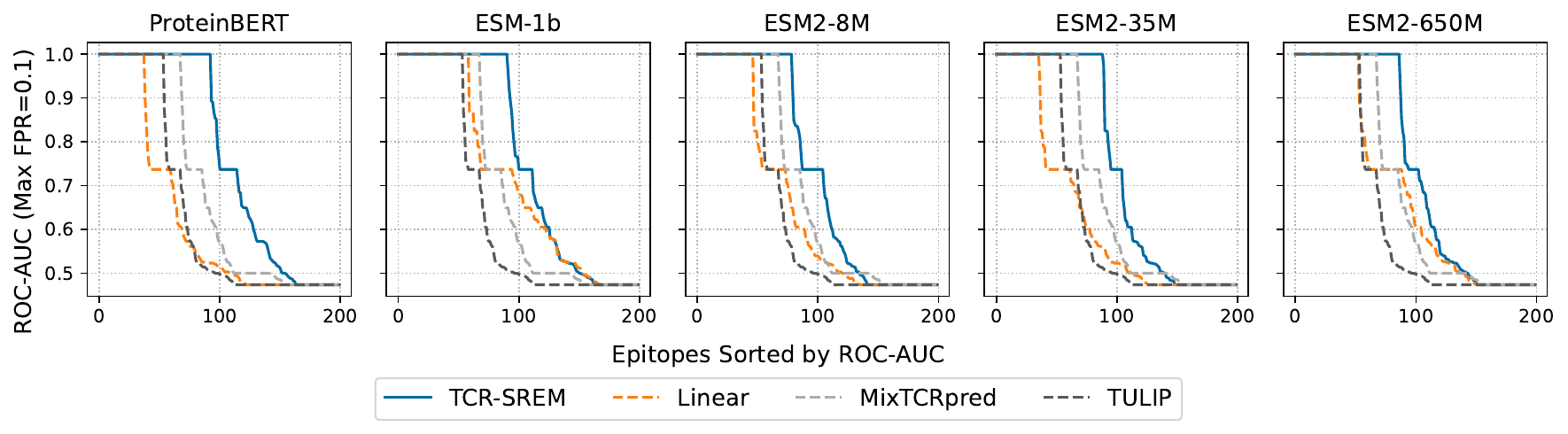}
    \caption{The epitope-wise ROC-AUC (FPR $<$ 0.1) evaluated on \ModelName[abbr] and baseline models. \ModelName[abbr] consistently achieves higher predictive ROC-AUC for more epitopes comparing against MixTCRpred and TULIP, where $\sim 20\%$ epitopes can achieve above $0.8$.}
    \label{fig:roc_auc_epiwise}
\end{figure*}

As shown in the Figure~\ref{fig:roc_auc_epiwise}, we demonstrate the epitope-wise ROC-AUC (FPR $< 0.1$) for all epitopes. Our \ModelName[abbr] achieves strong performance across a wide range of epitopes, with approximately 20\% reaching ROC-AUC values above 0.8, which shows superior generalization ability and predictive performance.

\subsection{Epitope-wise Precision, Recall, and F1 Analysis}
Considering that the test set is imbalanced, with substantially more negative than positive samples, we additionally report precision, recall, and F1 scores. Because these metrics require deterministic prediction labels, we apply a threshold of 0.5 for positive classification.
As shown in Table~\ref{tab:prf1}, models with structure regularization consistently achieve higher recall, reaching values above 0.84. In contrast, models without structure regularization obtain higher precision but suffer from significantly reduced recall, with decreases of approximately 10-20\%.
\begin{table}[ht]
    \centering
    \setlength{\tabcolsep}{6pt}
    \begin{tabular}{llcccc}
        \toprule
        \textbf{PLM} & \textbf{Metric} & \multicolumn{4}{c}{Top-$k$ @ Threshold = 0.5} \\
        \textbf{Backbone} & & \textbf{Top-100} & \textbf{Top-150} & \textbf{Top-200} & \textbf{Top-250} \\
        \midrule

        \multirow{3}{*}{ESM-1b}
            & Precision & 0.634 & 0.382 & 0.349 & 0.342 \\
            & Recall    & 0.891 & 0.544 & 0.482 & 0.472 \\
            & F1        & 0.741 & 0.449 & 0.405 & 0.397 \\
        \midrule

        \multirow{3}{*}{ESM2-8M}
            & Precision & 0.522 & 0.306 & 0.270 & 0.266 \\
            & Recall    & 0.877 & 0.588 & 0.512 & 0.506 \\
            & F1        & 0.655 & 0.403 & 0.354 & 0.349 \\
        \midrule

        \multirow{3}{*}{ESM2-35M}
            & Precision & 0.589 & 0.307 & 0.305 & 0.300 \\
            & Recall    & 0.844 & 0.410 & 0.406 & 0.396 \\
            & F1        & 0.694 & 0.351 & 0.348 & 0.342 \\
        \midrule

        \multirow{3}{*}{ESM2-650M}
            & Precision & 0.594 & 0.324 & 0.316 & 0.310 \\
            & Recall    & 0.887 & 0.441 & 0.426 & 0.415 \\
            & F1        & 0.712 & 0.373 & 0.363 & 0.355 \\
        \midrule

        \multirow{3}{*}{ProteinBERT}
            & Precision & 0.705 & 0.566 & 0.382 & 0.371 \\
            & Recall    & 0.925 & 0.536 & 0.380 & 0.369 \\
            & F1        & 0.800 & 0.550 & 0.381 & 0.370 \\
        \bottomrule
    \end{tabular}

    \caption{Precision, recall, and F1 of structure-regularized \ModelName[abbr] (threshold 0.5) across the Top-100, 150, 200, and 250 peptides, for each PLM backbone.}
    \label{tab:prf1}
\end{table}

\subsection{Extended Evaluation on Independent Comprehensive Benchmark}
\label{apdx:compb}
\begin{table}[ht]
    \centering
    \caption{AUPRC and ROC-AUC comparison of TCR-epitope binding prediction models on the Comprehensive benchmark. Our models based on different PLM backbones achieve competitive performance, with \ModelName[abbr] (ESM2-150M) and vibtcr obtaining the highest AUPRC, and \ModelName[abbr] (ProteinBERT)-1R (\ModelName[abbr] (ProteinBERT) with one epoch regularization) achieving the best ROC-AUC.}
    \label{tab:assess:auprc:full}
    \setlength{\tabcolsep}{3.5pt}
    \begin{tabular}{lll|lll}
    \toprule
    \textbf{Models} & \textbf{AUPRC} & \textbf{ROC-AUC} & \textbf{Models} & \textbf{AUPRC} & \textbf{ROC-AUC} \\
    \midrule
    \ModelName[abbr] (ProteinBERT)-1R & 0.575 & \textbf{0.571} & \ModelName[abbr] (ESM2-35M) & 0.571 & 0.538 \\
    \ModelName[abbr] (ESM2-150M) & \textbf{0.587} & 0.564 & \ModelName[abbr] (ProteinBERT)-0R & 0.509 & 0.532 \\
    PISTE & 0.567 & 0.564 & epiTCR & 0.499 & 0.527 \\
    vibtcr & \textbf{0.588} & 0.563 & pMTnet & 0.513 & 0.524 \\
    \ModelName[abbr] (ProteinBERT) & 0.550 & 0.541 & TCRconv & 0.548 & 0.519 \\
    MixTCRpred & 0.540 & 0.546 & \ModelName[abbr] (ESM2-650M) & 0.544 & 0.505 \\
    ERGOII & 0.540 & 0.546 & \ModelName[abbr] (ESM2-8M) & 0.502 & 0.475 \\
    \ModelName[abbr] (ESM1b) & 0.557 & 0.540 & & & \\
    \bottomrule
    \end{tabular}
\end{table}
We compare \ModelName[abbr] with different PLM backbones and more models including ERGOII~\cite{springer2021contribution}, epiTCR~\cite{pham2023epitcr}, TCRconv~\cite{jokinen2023tcrconv}, and pMTnet~\cite{lu2021deep} in Table~\ref{tab:assess:auprc:full}.
Notably, using the same backbone with different numbers of structure regularization epochs leads to distinct performance. Without structure regularization, the AUPRC drops to 0.509; with one epoch, it improves to 0.575; however, with three epochs, it decreases to 0.550. This indicates that generalization is sensitive to the extent of structure regularization. A limited number of regularization epochs helps guide the model to learn meaningful contact patterns in TCR-peptide binding, whereas excessive regularization degrades generalization performance.

\subsection{IMMREP23 ROC-AUC Analysis}

To further validate our findings, we evaluated all models on the IMMREP23 benchmark~\cite{nielsen2024lessons} and compared their performance against MixTCRpred. All models were trained using our compiled training dataset. The ESM-1b backbone achieves the best overall performance (ROC-AUC $@$ FPR$<$0.1 = 0.602), whereas ProteinBERT performs the worst (ROC-AUC $@$ FPR$<$0.1 = 0.546). Notably, all PLM backbones integrated with \ModelName[abbr] consistently outperform MixTCRpred (ROC-AUC $@$ FPR$<$0.1 = 0.482). This performance difference can be partially attributed to the modeling focus of the respective backbones: ESM-1b places greater emphasis on CDR3a-peptide interactions, while ProteinBERT primarily captures CDR3b-peptide interactions. In contrast, the relatively weaker performance of all these models may stem from several factors, including the presence of non-human and non-murine samples in our training dataset, the absence of epitope-specific optimization for the IMMREP23 task. For MixTCRpred, the exclusion of samples lacking complete CDR regions is the other important reason.

\begin{table}[ht]
    \centering
    \caption{Comparison of ROC-AUC scores with the false positive rate restricted to 0.1 on IMMREP23 benchmark for \ModelName[abbr] with various PLM backbones and MixTCRpred trained on our training dataset.}
    \label{tab:immrep23}
    \begin{tabular}{lcccccc}
        \toprule
        Backbones &  ProteinBERT & ESM-1b & ESM2-8M & ESM2-35M & ESM2-650M & MixTCRpred\\
        \midrule
        ROC-AUC &  0.546 & 0.602 & 0.575 & 0.572 & 0.577 & 0.482 \\
        \bottomrule
    \end{tabular}
    
\end{table}

\section{Impact of Generated Structures}
In this section, we first describe the preparation of predicted structures. We then provide extended analysis showing that real-structure regularization significantly improves performance over predicted structures, and report detailed per-sample pairwise normalized RMSD to illustrate the binding-site diversity difference between real and predicted structures.

\subsection{Preparation of Predicted Structure}
To prepare the predicted-structure dataset, we first extracted sequences from the TCR-XAI regularization dataset. Each entry includes TCR $\alpha$, TCR $\beta$, peptide, and MHC sequences (a two-chain MHC is recorded as two separate chains). For each method, we define a proxy resolution that rescales the model's per-residue confidence to a range comparable to experimental resolution:
\[
r = \frac{100 - \bar{\phi}_{\min}}{10}, \qquad
\bar{\phi}_{\min} = \min\!\left(\frac{\sum\phi_a}{|\phi_a|},\, \frac{\sum\phi_b}{|\phi_b|},\, \frac{\sum\phi_e}{|\phi_e|}\right),
\]
where $\phi_a$, $\phi_b$, and $\phi_e$ are the per-residue confidence scores (pLDDT or equivalent) of the CDR3a, CDR3b, and peptide chains, and $\bar{\phi}_{\min}$ is the lowest of the three chain-averaged confidences.

\paragraph{AlphaFold3.} We used the AlphaFold3 server (\url{https://alphafoldserver.com/}), treating each component of the TCR-epitope complex as an independent chain, and took the rank-1 model as the predicted structure. Proxy resolution follows the equation above with $\phi$ taken as pLDDT.

\paragraph{tFold-TCR.} We deployed a local instance following \url{https://github.com/TencentAI4S/tfold} and ran inference on CPU. Here confidence is the B-factor (or equivalent) reported in the output PDB, normalized to $[0,1]$, so proxy resolution is computed as
\[
r = 10\left(1 - \bar{\phi}_{\min}\right).
\]

\paragraph{TCRModel2.} We used the web service (\url{https://tcrmodel.ibbr.umd.edu/}) with $\phi$ taken as pLDDT. For MHC-I, the server accepts only a single chain, so we omitted the $\beta_2$-microglobulin sequence when present; for MHC-II, the server truncates sequences longer than 11 residues, and we padded the truncated positions with the nearest values when computing contact maps. Three structures were omitted due to server errors: 4Z7W and 4OZH returned an unknown error, and 5KSB returned ``cannot find core.''

\subsection{Test of Significant for Impact of Epitope-wise ROC-AUC}
Because the improvement between regularizing with real versus AlphaFold3-predicted structures is only 1.2\%, we report a significance test in Table~\ref{tab:roc_auc:generated_structure:p}. All $p$-values are below 0.005, confirming that the improvements are statistically significant.
\begin{table}[ht]
    \centering
    \caption{ROC-AUC scores (FPR restricted to 0.1) for \ModelName[abbr] regularized with predicted structures, across the Top-100, 150, 200, 250, and 300 peptides. Significance versus real-structure regularization shown in parentheses ($p$-value).}
    \label{tab:roc_auc:generated_structure:p}
    \begin{tabular}{llllll}
        \toprule
        \multirow{2}{*}{\textbf{Structures}} & \multicolumn{5}{c}{Top-$k$ ROC-AUC @ FPR$\le$0.1} \\
         & \textbf{100} & \textbf{150} & \textbf{200} & \textbf{250} & \textbf{300} \\
        \midrule
        Real & \textbf{0.953} & \textbf{0.821} & \textbf{0.734} & \textbf{0.682} & \textbf{0.655}\\
        AlphaFold3 & 0.941 (2.7$e-$4) & 0.804 (1.2$e-$10) & 0.721 (2.4$e-$10) & 0.672 (3.6$e-$10)	& 0.646  (4.4$e-$10) \\
        TCRModel2 & 0.929 (1.7$e-$7) & 0.790 (5.5$e-$14) & 0.711 (2.2$e-$13) & 0.663 (4.6$e-$13)	& 0.638  (6.7$e-$13) \\
        tFold-TCR & 0.913 (2.4$e-$6) & 0.778 (1.6$e-$13) & 0.702 (5.7$e-$13) & 0.656 (1.1$e-$12) & 0.632  (1.6$e-$12) \\
        \bottomrule
    \end{tabular}
\end{table}

\begin{figure}[hbt]
    \centering
    \includegraphics[width=\linewidth]{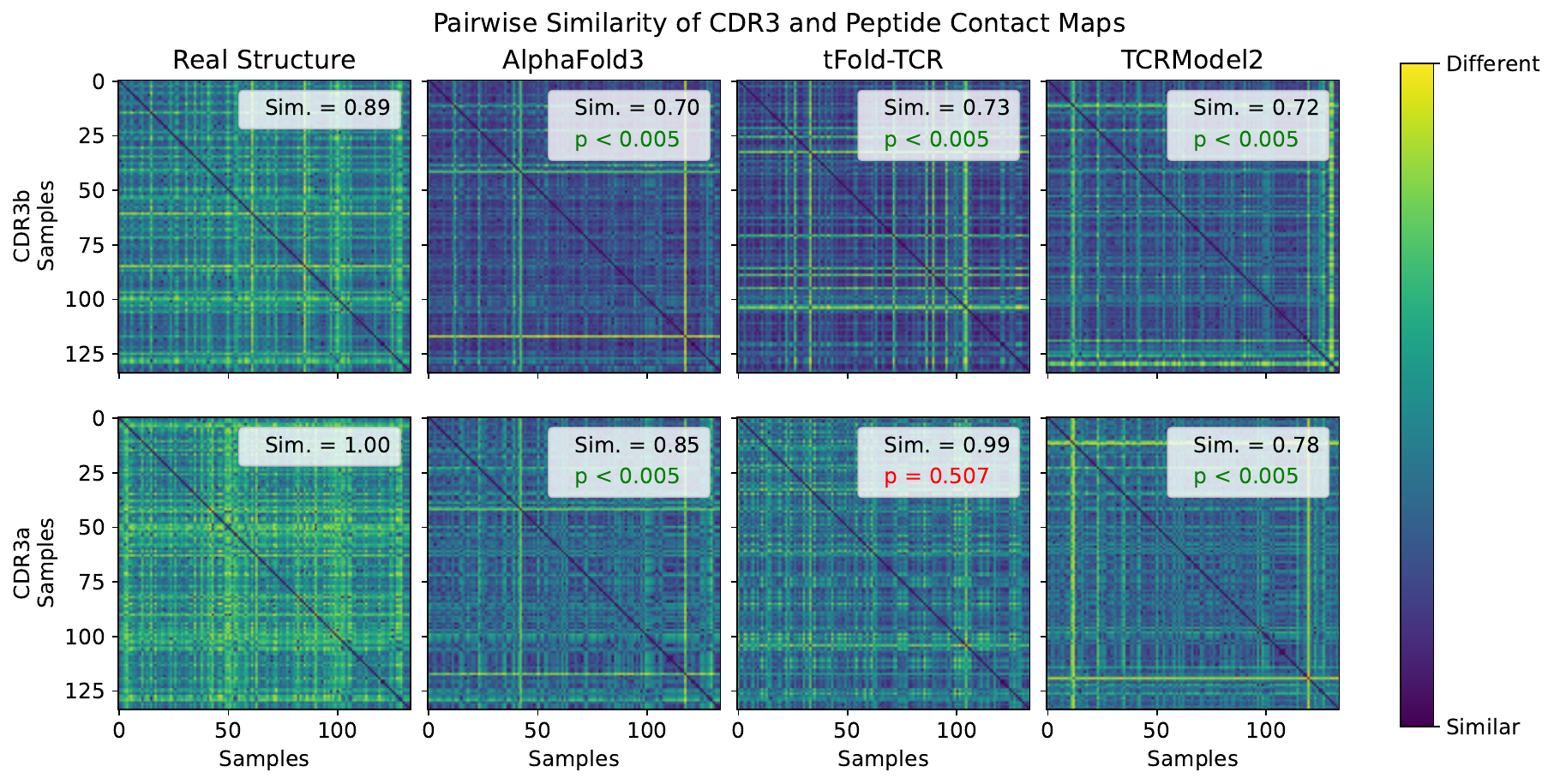}
    \caption{Pairwise RMSD of CDR3 and Peptide Contact Maps}
    \label{fig:pairwise_rmsd}
\end{figure}
\subsection{Binding-Site Diversity Differs between Predicted and Real Structures}
\label{apdx:sdiver}
As shown in Figure~\ref{fig:pairwise_rmsd}, we compute pairwise similarity of the CDR-peptide distance matrices and use averaged similarity to indicate binding site diversity, where the pairwise similarity is RMSD between any pair of z-scored contact distance matrices. Except CDR3a-peptide predicted by tFold-TCR, for all other predicted structures, experimentally resolved structures showing significant higher contact site diversity with $p<0.005$.

\section{Ablation Studies}
\label{apdx:ablation}
To investigate the effectiveness of structural regularization and the sensitivity of \ModelName[abbr] to key hyperparameters, we conducted ablation studies using the ESM2-8M backbone, which outperforms baseline models while remaining efficient enough to allow testing of many hyperparameter combinations. Specifically, we examined four factors: (1) \emph{structural regularization}, evaluating its impact on predictive performance and interpretation quality, as well as the relationship between the number of available structures and model performance; (2) \emph{mixing strategy}, comparing maximum versus average pooling for aggregating CDR3a-peptide and CDR3b-peptide contact prototype scores; (3) \emph{regularization threshold}, assessing the effect of varying the threshold $\epsilon$ from 0.6 to 0.9 on structural regularization strength; and (4) \emph{regularization frequency}, varying the number of structural regularization epochs applied after each standard training epoch.

\subsection{Structure Regularization}
\begin{figure}[th]
    \centering
    \includegraphics[width=\linewidth]{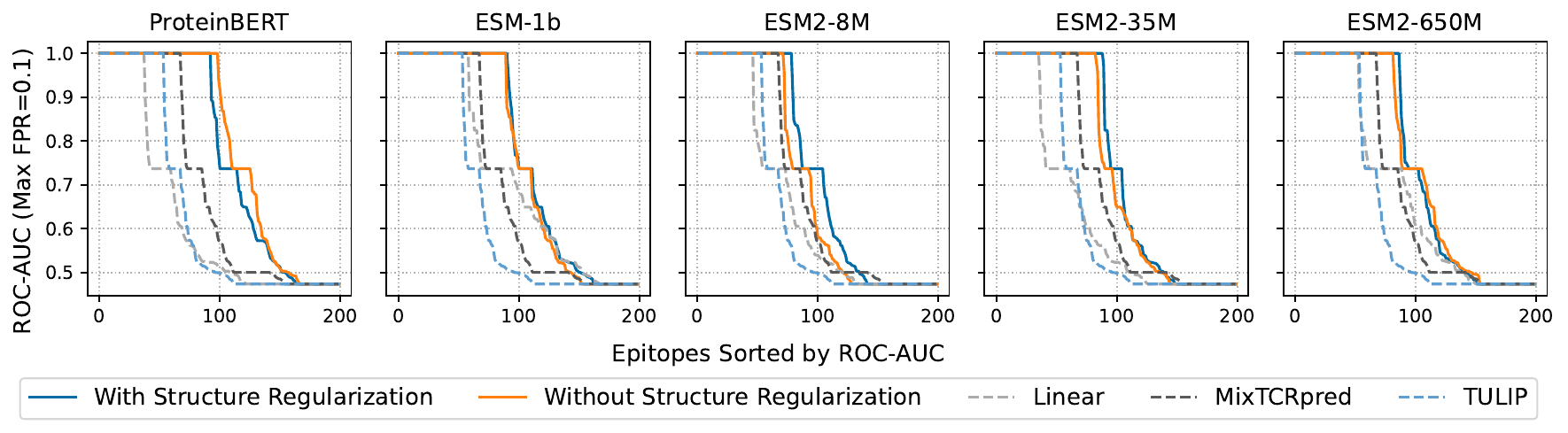}
    \caption{The epitope-wise ROC-AUC (Max FPR = 0.1) evaluated on \ModelName[abbr] with or without structure regularization and baseline models.}
    \label{fig:roc_auc_epiwise:ablation}
\end{figure}
To evaluate the impact of structure regularization, we conduct ablation studies to assess its effect on both performance and interpretability. In the absence of structure regularization, the model learns the contact prototypes in an unsupervised manner and infers contact prototypes of the CDR3-peptide based on only sequence inputs. In practice, this was implemented by omitting the structure regularization epochs following the standard training epochs.
Overall, we find that structure regularization yields minor improvements, while interpretability has major, statistically significant, improvements.

\paragraph{Impact of Regularization on Predictive Performance:} 
\begin{table}[th]
    \centering
    \caption{Comparison of ROC-AUC scores with the false positive rate restricted to 0.1 across the Top-100, Top-150, Top-200, Top-250, and Top-300 peptides among \ModelName[abbr] with or without structure regularization.}
    \label{tab:roc_auc:ablation}
    \begin{tabular}{lllllll}
        \toprule
        \multirow{2}{*}{\textbf{Model}} & \multirow{2}{*}{\textbf{Embedding}} & \multicolumn{5}{c}{Top-$k$ ROC-AUC @ FPR$\le$0.1} \\
         & & \textbf{100} & \textbf{150} & \textbf{200} & \textbf{250} & \textbf{300} \\
        \midrule
        \multirow{3}{*}{Omitting} & ESM-1b & 0.982 & 0.854 & 0.759 & 0.702 & 0.672 \\
        \multirow{3}{*}{Structure} & ProteinBERT & \textbf{0.999} & \textbf{0.895} & \textbf{0.792} & \textbf{0.728} & \textbf{0.695} \\
        \multirow{3}{*}{Regularization} & ESM2-8M & 0.926 & 0.786 & 0.708 & 0.661 & 0.636 \\
          & ESM2-35M & 0.960 & 0.823 & 0.735 & 0.683 & 0.655 \\
         & ESM2-650M & 0.960 & 0.836 & 0.746 & 0.691 & 0.663 \\
         
        \hline
        \multirow{4}{*}{Structure} & ESM-1b & \textbf{0.986} & \textbf{0.862} & \textbf{0.766} & \textbf{0.707} & \textbf{0.677} \\
        \multirow{4}{*}{Regularization} & ProteinBERT & 0.989 & 0.871 & 0.773 & 0.713 & 0.681 \\
        & ESM2-8M & \textbf{0.953} & \textbf{0.821} & \textbf{0.734} & \textbf{0.682} & \textbf{0.655} \\
        & ESM2-35M & \textbf{0.977} & \textbf{0.838} & \textbf{0.747} & \textbf{0.692} & \textbf{0.663} \\
         & ESM2-650M & \textbf{0.971} & \textbf{0.837} & \textbf{0.746} & \textbf{0.692} & \textbf{0.663} \\
         
        \bottomrule
    \end{tabular}
\end{table}

\begin{table}[ht]
    \centering
    \caption{Precision, recall, and F1 (threshold $=0.5$) across Top-100/150/200/250 peptides.
    For each PLM backbone, rows compare \ModelName[abbr] without ($-$) and with ($+$) structure regularization.}
    \setlength{\tabcolsep}{3pt}
    \begin{tabular}{ll *{12}{c}}
        \toprule
        & & \multicolumn{4}{c}{\textbf{Precision @ 0.5}}
          & \multicolumn{4}{c}{\textbf{Recall @ 0.5}}
          & \multicolumn{4}{c}{\textbf{F1 @ 0.5}} \\
        \cmidrule(lr){3-6}\cmidrule(lr){7-10}\cmidrule(lr){11-14}
        \textbf{PLM Backbone} & \textbf{Structure}
          & \textbf{T100} & \textbf{T150} & \textbf{T200} & \textbf{T250}
          & \textbf{T100} & \textbf{T150} & \textbf{T200} & \textbf{T250}
          & \textbf{T100} & \textbf{T150} & \textbf{T200} & \textbf{T250} \\
        \midrule

        \multirow{2}{*}{ESM-1b}
          & $-$ & 0.836 & 0.498 & 0.481 & 0.467 & 0.631 & 0.184 & 0.174 & 0.167 & 0.719 & 0.269 & 0.255 & 0.246 \\
          & $+$ & 0.634 & 0.382 & 0.349 & 0.342 & 0.891 & 0.544 & 0.482 & 0.472 & 0.741 & 0.449 & 0.405 & 0.397 \\
        \addlinespace

        \multirow{2}{*}{ESM2-8M}
          & $-$ & 0.729 & 0.477 & 0.464 & 0.373 & 0.656 & 0.269 & 0.256 & 0.219 & 0.691 & 0.344 & 0.330 & 0.276 \\
          & $+$ & 0.522 & 0.306 & 0.270 & 0.266 & 0.877 & 0.588 & 0.512 & 0.506 & 0.655 & 0.403 & 0.354 & 0.349 \\
        \addlinespace

        \multirow{2}{*}{ESM2-35M}
          & $-$ & 0.764 & 0.468 & 0.449 & 0.434 & 0.723 & 0.269 & 0.256 & 0.247 & 0.743 & 0.342 & 0.326 & 0.315 \\
          & $+$ & 0.589 & 0.307 & 0.305 & 0.300 & 0.844 & 0.410 & 0.406 & 0.396 & 0.694 & 0.351 & 0.348 & 0.342 \\
        \addlinespace

        \multirow{2}{*}{ESM2-650M}
          & $-$ & 0.827 & 0.513 & 0.486 & 0.480 & 0.613 & 0.219 & 0.195 & 0.187 & 0.705 & 0.307 & 0.278 & 0.269 \\
          & $+$ & 0.594 & 0.324 & 0.316 & 0.310 & 0.887 & 0.441 & 0.426 & 0.415 & 0.712 & 0.373 & 0.363 & 0.355 \\
        \addlinespace

        \multirow{2}{*}{ProteinBERT}
          & $-$ & 0.931 & 0.730 & 0.537 & 0.524 & 0.736 & 0.382 & 0.199 & 0.190 & 0.822 & 0.502 & 0.290 & 0.279 \\
          & $+$ & 0.705 & 0.566 & 0.382 & 0.371 & 0.925 & 0.536 & 0.380 & 0.369 & 0.800 & 0.550 & 0.381 & 0.370 \\
        \bottomrule
    \end{tabular}
    
    \label{tab:prf1:ablation:full}
\end{table}

Table~\ref{tab:roc_auc:ablation} and Figure~\ref{fig:roc_auc_epiwise:ablation} shows results of our \ModelName[abbr] with different backbones with and without  structure regularization. We see that regularization generally leads to a performance increase across most PLM backbones. Notably, ESM2-8M, a smaller PLM, experiences an increase of approximately 2.5\% in ROC-AUC, while larger models experience smaller improvements.
ProteinBERT was an exception, exhibiting a modest improvement without structure regularization (of about 1\%), which we attribute to its strong pretraining on UniProtKB/UniRef90 ($\sim$ 106 million protein sequences) and with gene ontology (GO) annotation prediction task that empowers it to produce high-quality protein features. We also provide epitope-wise precision, recall, and F1 for reference in Table~\ref{tab:prf1:ablation:full}.

\begin{table}[ht]
\centering
    \caption{Comparison of structure regularization and non-regularization on BRHR across different protein language model backbones, where $a \to b$ denotes the evaluation of residues in chain $b$ with respect to chain $a$.}
    \label{tab:brhr:ablation}
    \begin{tabular}{llcccc}
        \toprule
        \multirow{3}{*}{\textbf{Embeddings}} & \textbf{Structure} & \textbf{Peptide} & \textbf{Peptide} & \textbf{CDR3a} & \textbf{CDR3b} \\
        & (-) without & $\downarrow$ & $\downarrow$ & $\downarrow$ & $\downarrow$ \\
        & (+) with  & \textbf{CDR3a} & \textbf{CDR3b} & \textbf{Peptide} & \textbf{Peptide} \\
        \midrule
        \multirow{2}{*}{ProteinBERT} 
        & (+) & 0.568 & 0.996 & 0.392 & 0.855 \\
        & (-) & 0.842 & 0.848 & 0.786 & 0.743 \\
        \hline
        \multirow{2}{*}{ESM-1b} 
        & (+) & 0.833 & 0.944 & 0.818 & 0.852 \\
        & (-) & 0.876 & 0.854 & 0.817 & 0.788 \\
        \hline
        \multirow{2}{*}{ESM2-8M} 
        & (+) & 0.604 & 0.801 & 0.804 & 0.816 \\
        & (-) & 0.842 & 0.840 & 0.784 & 0.779 \\
        \hline
        \multirow{2}{*}{ESM2-35M} 
        & (+) & 0.834 & 0.944 & 0.782 & 0.746 \\
        & (-) & 0.847 & 0.857 & 0.687 & 0.773 \\
        \hline
        \multirow{2}{*}{ESM2-650M} 
        & (+) & 0.961 & 0.820 & 0.788 & 0.809 \\
        & (-) & 0.861 & 0.823 & 0.776 & 0.789 \\
        \bottomrule
    \end{tabular}
    
\end{table}

\paragraph{Impact of Regularization on Interpretation Quality:} To evaluate the impact of structure regularization on the quality of contact prototype interpretations, we evaluated the BRHR of \ModelName[abbr] with and without structure regularization. As shown in Table~\ref{tab:brhr:ablation}, structure regularization leads to a substantial increase in BRHR for CDR3b-peptide interactions, with ProteinBERT showing an increase of approximately 10\% (p-value $< 0.001$ compared to all baselines). Interestingly, in the absence of structure regularization, the model assigns more balanced weights to CDR3a-peptide and CDR3b-peptide interactions, rather than focusing predominantly on a single interaction as observed in \ModelName[abbr] with ProteinBERT.

\paragraph{Impact of Regularization on Contact Prototype:}
\begin{figure}[hbt]
    \centering
    \includegraphics[width=0.5\linewidth]{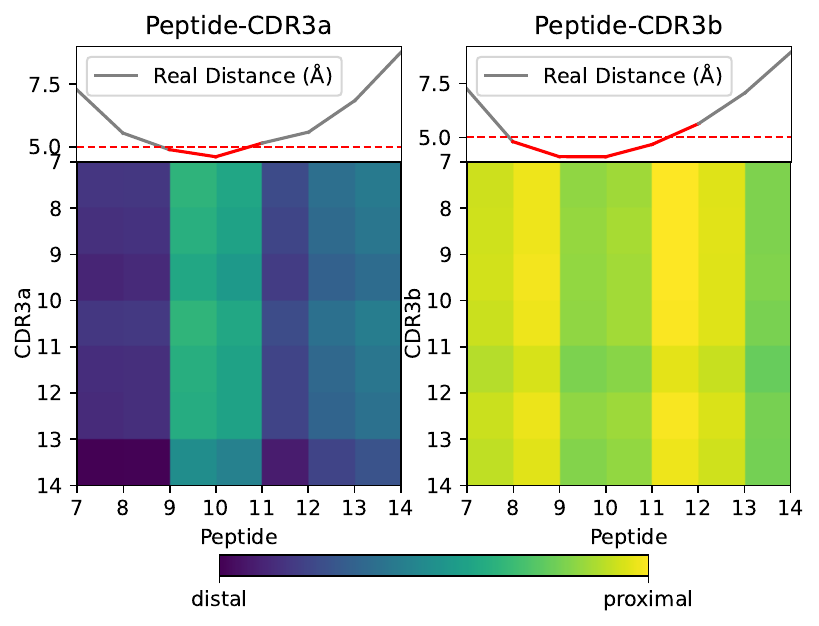}
    \caption{Average centralized contact prototypes on the TCR-XAI benchmark in the absence of structure regularization.}
    \label{fig:ablation:distpb}
\end{figure}
We also visualized the contact prototypes of \ModelName[abbr] with ProteinBERT, omitting structure regularization, on the TCR-XAI benchmark. As shown in Figure~\ref{fig:ablation:distpb}, the model exhibits reasonably balanced accuracy for both CDR3a-peptide and CDR3b-peptide interactions, rather than focusing predominantly on CDR3b-peptide. However, the resolution and quality of the CDR3b-peptide interaction remain substantially lower than in the structure-regularized models. This proves our observation from quantitative BRHR comparison between structure regularization and omitting structure regularization.

\begin{figure}[ht]
    \centering
    \includegraphics[width=\linewidth]{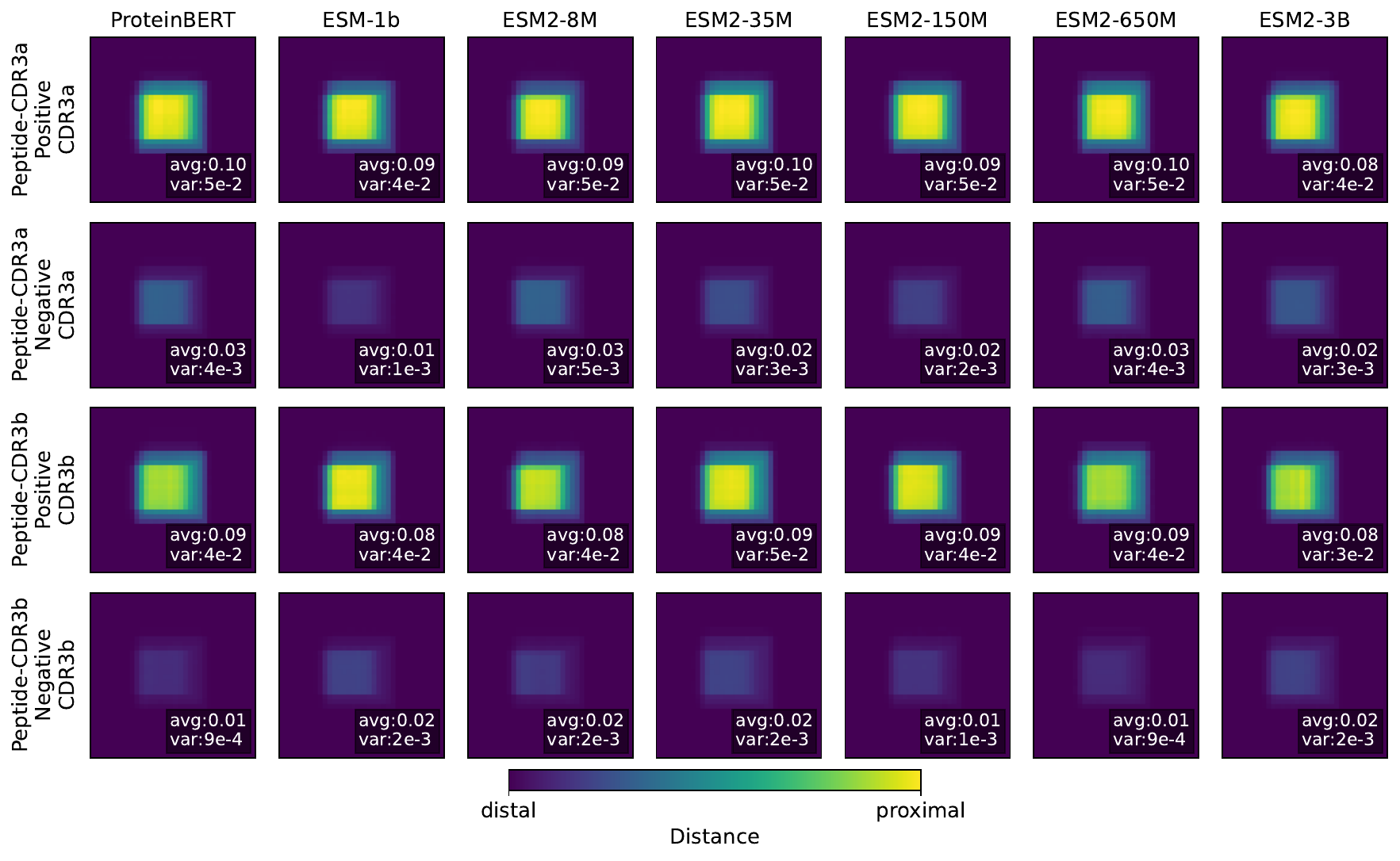}
    \caption{Average centralized contact prototypes on the training dataset in the absence of structure regularization from \ModelName[abbr] with various PLM backbones.}
    \label{fig:ablation:distall}
\end{figure}

Finally, we visualized the contact prototypes across all backbones on the entire training dataset. As shown in Figure~\ref{fig:ablation:distall}, for \ModelName[abbr] without structure regularization, the variance of contact prototypes from positive samples is substantially higher than that of negative samples, indicating that the model can still learn meaningful contact patterns from sequence data alone. Nonetheless, structure regularization ensures the learning of higher-quality prototypes.

\subsection{Regularization Set Size}
\begin{figure}[ht]
    \centering
    \includegraphics[width=0.5\linewidth]{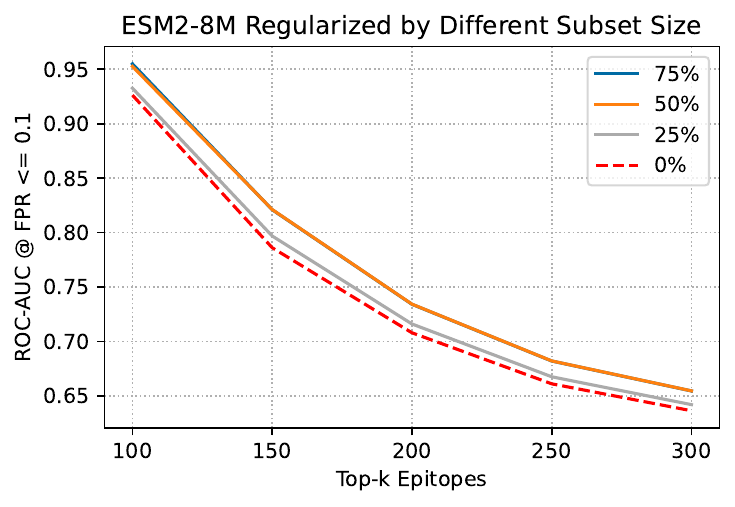}
    \caption{Comparison of ROC-AUC scores with the false positive rate restricted to 0.1 across \ModelName[abbr] with ESM2-8M regularized by different TCR-XAI subset size including 75\%, 50\%, and 25\%. Structural regularization consistently improves generalization performance across all subset sizes monotonically.}
    \label{fig:diffstruct}
\end{figure}
We also examined the effect of the number of structures used for regularization on model performance. We regularized \ModelName[abbr] with the ESM2-8M backbone using randomly selected subsets of TCR-XAI benchmark structures, comprising 25\%, 50\%, and 75\% of the full set. As shown in Figure~\ref{fig:diffstruct}, structural regularization consistently improves generalization performance across all subset sizes monotonically (0.5\%, 2.5\%, and 2.5\% improvement, respectively). We note that, interestingly, even using 25\% of the structures used (68 complexes) yields a minor improvement.

\subsection{Mixing Strategy and Regularization Thresholds}
To examine how different mixing strategies of CDR3a-peptide and CDR3b-peptide contact scores, as well as regularization thresholds, influence model performance, we conduct ablation studies varying both factors. The mixing strategies include maximum (Max.) and average pooling (Avg.), and the thresholds ($\epsilon$) range from 0.6 to 0.9.
As shown in Table~\ref{tab:ablation:maxavgthreshold}, the choice of mixing method has a minor impact, with performance differences within 5\%. In contrast, the regularization threshold has a pronounced effect: the best results are achieved at $\epsilon = 0.75-0.8$, while larger or smaller values reduce performance by approximately 10\%.
\begin{table}[ht]
    \centering
    \caption{Ablation study on regularization thresholds ($\epsilon$) for max and average pooling, evaluated by ROC-AUC (Max FPR = 0.1) across different top-$k$ peptide subsets.}
    \label{tab:ablation:maxavgthreshold}
    \begin{tabular}{lllllll}
        \toprule
        Regularize & Mix & \multicolumn{5}{c}{Top-$k$ ROC-AUC @ FPR $\le$ 0.1} \\
        Threshold & Method & \textbf{100} & \textbf{150} & \textbf{200} & \textbf{250} & \textbf{300} \\
        \midrule
        \multirow{2}{*}{$\epsilon=0.6$} & Max. & 0.916 & 0.785 & 0.707 & 0.661 & 0.636 \\
         & Avg. & 0.896 & 0.763 & 0.691 & 0.647 & 0.624 \\
        \hline
        \multirow{2}{*}{$\epsilon=0.7$} & Max. & 0.844 & 0.722 & 0.660 & 0.623 & 0.603 \\
         & Avg. & 0.893 & 0.763 & 0.690 & 0.647 & 0.624 \\
         \hline
        \multirow{2}{*}{$\epsilon=0.75$} & Max. & 0.932 & 0.794 & 0.714 & 0.666 & 0.641 \\
         & Avg. & 0.961 & 0.825 & 0.737 & 0.684 & 0.657 \\
        \hline
        \multirow{2}{*}{$\epsilon=0.8$} & Max. & 0.953 & 0.821 & 0.734 & 0.682 & 0.655 \\
         & Avg. & 0.948 & 0.811 & 0.726 & 0.676 & 0.649 \\
        \hline
        \multirow{2}{*}{$\epsilon=0.85$} & Max. & 0.881 & 0.749 & 0.680 & 0.639 & 0.617 \\
         & Avg. & 0.897 & 0.766 & 0.693 & 0.649 & 0.626 \\
        \hline
        \multirow{2}{*}{$\epsilon=0.9$} & Max. & 0.878 & 0.749 & 0.681 & 0.639 & 0.617 \\
         & Avg. & 0.922 & 0.786 & 0.708 & 0.661 & 0.636 \\
        \bottomrule
    \end{tabular}
\end{table}

\subsection{Regularization Frequency}
\begin{table}[ht]
    \centering
    \caption{Ablation study on regularization frequency evaluated by ROC-AUC (Max FPR = 0.1) across different top-$k$ peptide subsets.}
    \label{tab:ablation:epoch}
    \begin{tabular}{llllll}
        \toprule
        Regularize & \multicolumn{5}{c}{Top-$k$ ROC-AUC @ FPR $\le$ 0.1} \\
        Epoch & \textbf{100} & \textbf{150} & \textbf{200} & \textbf{250} & \textbf{300} \\
        \midrule
        1 & 0.899 & 0.762 & 0.690 & 0.647 & 0.624 \\
        3 & 0.973 & 0.839 & 0.748 & 0.693 & 0.664 \\
        5 & 0.953 & 0.821 & 0.734 & 0.682 & 0.655 \\
        7 & 0.908 & 0.774 & 0.699 & 0.654 & 0.630 \\
        9 & 0.850 & 0.728 & 0.664 & 0.626 & 0.606 \\
        \bottomrule
    \end{tabular}
\end{table}
Another key factor influencing structure regularization is the regularization frequency, defined as the number of structure regularization epochs applied after each standard training epoch. We evaluated the impact of varying this frequency from 1 to 9 epochs.
As reported in Table~\ref{tab:ablation:epoch}, the model achieves optimal performance, ROC-AUC exceeding 0.95, when the number of regularization epochs is 3 or 5. Both higher and lower frequencies lead to a substantial performance drop of over 5\%.

\section{Availability and Implementation}
All experiments were performed on three Ubuntu servers equipped with two NVIDIA A2000, one NVIDIA RTX3090, and two NVIDIA 6000 Ada GPUs respectively. To facilitate efficient training and evaluation with large protein language models (PLMs), we first extracted and de-duplicated amino acid sequences from CDR3a, CDR3b, and peptides. The corresponding embeddings were pre-computed using PLMs and stored for downstream use. During model training and evaluation, sequence indices were used to retrieve and assemble these pre-computed representations. Each model was trained for 150 epochs with a batch size of 512 and a learning rate of $1 \times 10^{-3}$ using the AdamW optimizer. A dropout rate of $0.2$ was applied to enhance generalization performance.

\section{Data \& Code Availability:}
The code and data used in this paper can be found at GitHub repository: \url{https://github.com/Tulane-Mettu-Landry-Lab/tcr-sr}.


\end{document}